\documentclass[iop, twocolumn, tighten]{emulateapj}
\usepackage{amsmath}
\usepackage{color}
\definecolor{redgray}{rgb}{0.4,0.0,0.0}
\definecolor{red}{rgb}{0.0,0.0,0.0}
\definecolor{blue}{rgb}{0.0,0.0,0.8}
\definecolor{green}{rgb}{0.0,0.5,0.0}
\definecolor{purple}{rgb}{0.5,0.0,0.8}
\usepackage{ulem}

\newcommand{\fif}{\mathit{f}}

\slugcomment{}
\shortauthors{}
\shorttitle{}
\begin{document}
\title{A Linear and Quadratic Time-Frequency Analysis of Gravitational Waves from Core-Collapse Supernovae}

\author{Hajime Kawahara\altaffilmark{1,2}, Takami Kuroda\altaffilmark{3}, Tomoya Takiwaki\altaffilmark{4}, Kazuhiro Hayama\altaffilmark{5}, and Kei Kotake\altaffilmark{5}} 
\altaffiltext{1}{Department of Earth and Planetary Science, 
The University of Tokyo, Tokyo 113-0033, Japan}
\altaffiltext{2}{Research Center for the Early Universe, 
School of Science, The University of Tokyo, Tokyo 113-0033, Japan}
\altaffiltext{3}{Institut f{\"u}r Kernphysik, Technische Universit{\"a}t Darmstadt, Schlossgartenstrasse 9, D-64289 Darmstadt, Germany}
\altaffiltext{4}{Division of Theoretical Astronomy, National Astronomical Observatory of Japan, 2-21-1, Osawa, Mitaka, Tokyo 181-8588, Japan}
\altaffiltext{5}{Department of Applied Physics, Fukuoka University, 8-19-1, Jonan, Nanakuma, Fukuoka 814-0180, Japan}

\email{Electronic address: kawahara@eps.s.u-tokyo.ac.jp}

\begin{abstract}
  Recent core-collapse supernova (CCSN) simulations have predicted several distinct features in 
  gravitational-wave (GW) spectrograms, including a ramp-up signature due to the $g$-mode oscillation of the proto-neutron star (PNS) and an excess in the low-frequency domain (100$\sim$300 Hz) potentially induced by the standing accretion shock instability (SASI). These predictions motivated us to perform a sophisticated time-frequency analysis (TFA) of the GW signals, aimed at preparation for future observations. {\color{red} By reanalyzing a gravitational waveform obtained in a three-dimensional general-relativistic CCSN simulation, we show that both the spectrogram with an adequate window and the quadratic TFA separate the multimodal GW signatures much more clearly compared with the previous analysis.} We find that the observed low-frequency excess during the SASI active phase is divided into two components, a stronger one at 130 Hz and an overtone at 260 Hz, both of which evolve quasi-statically during the simulation time. We also identify a new mode whose frequency varies from 700 to 600 Hz. Furthermore, we develop the quadratic TFA for the Stokes $I,Q,U,$ and $V$ parameters as a new tool to investigate the GW circular polarization. We demonstrate that the polarization states that randomly change with time after bounce are associated with the PNS $g$-mode oscillation, whereas a slowly changing polarization state in the low-frequency domain is connected to the PNS core oscillation. This study demonstrates the capability of the sophisticated TFA for diagnosing the polarized CCSN GWs in order to explore their complex nature.

\end{abstract}
\keywords{}
\section{Introduction}
The first detections of gravitational waves (GWs) from the merging systems of black holes and neutron stars opened new avenues for GW astronomy (e.g., \citet{2016PhRvL.116f1102A,2017PhRvL.119p1101A}). The network of the three detectors (LIGO-Hanford/Livingston, VIRGO) has not only significantly improved the sky localization of the source, but also made the first probe into the GW polarization \citep{GW2_virgo} possible. The four-detector era will occur in the near future using KAGRA \citep{kagra17} and the nature of garden variety astrophysical sources (e.g., \cite{schutz09}), 
 including core-collapse supernovae (CCSNe, e.g., \citet{janka17} for a review), is expected to be determined.

 Unlike the GW signals from compact binary coalescence where a template-based search is best suited (e.g., \citet{rasio99}), 
 gravitational waveforms from CCSNe are of essentially stochastic nature. This is because the waveforms are all essentially affected by
turbulence in the postbounce core, which is governed by
multi-dimensional (multi-D) hydrodynamics.
In order to clarify the GW emission mechanisms, extensive numerical simulations have been performed 
in different contexts (e.g., \citet{Dimmelmeier08,Scheidegger10,CerdaDuran13,Ott13,Yakunin15,KurodaT14,pan18,viktoriya18} and
 \citet{Kotake13,Ott09} for a review). 
For canonical supernova progenitors \citep{Heger05}, the core rotation is generally
 too slow to affect the dynamics (e.g., \citet{takiwaki16,summa17}).
For such progenitors, the GW emission takes place 
 in the postbounce phase, 
 which is characterized by prompt convection, neutrino-driven 
convection, PNS convection, the standing accretion shock instability (SASI), and the $g$(/$f$)-mode 
oscillation of the PNS
 surface (e.g., \citet{EMuller97,EMuller04,Murphy09,Kotake09,BMuller13,viktoriya18}).
 
 The most distinct GW emission process commonly seen in recent 
self-consistent three-dimensional (3D) models 
 is that of the PNS oscillation \citep{2016ApJ...829L..14K,Andresen17,Yakunin17}.
 The characteristic GW frequency increases almost monotonically 
with time due to an accumulating accretion to the PNS, which ranges 
from $\sim 100$ to $1000$ Hz.
On the other hand, the typical 
frequency of the SASI-induced GW signals is concentrated in 
 the lower frequency range of $\sim 100$ to $250$ Hz and persists when the SASI 
dominates over neutrino-driven convection \citep{2016ApJ...829L..14K,Andresen17}.
 The detection of these distinct GW features could be the key to infer which is the most dominant 
 in the supernova engine, neutrino-driven 
convection, or the SASI \citep{Andresen17}.


To discuss the detectability of these signal predictions, previous studies have traditionally relied on a GW spectrogram analysis, aimed at specifying the GW feature by the excess power in the time-frequency domain (by taking the square norm of the short-time Fourier transform). However, it is well known that the spectrogram analysis has a trade-off relation between the time and frequency resolution, leading to a limited time-frequency localization. Because of this drawback, some features could have been 
potentially overlooked in the previous spectrogram analysis in the context of CCSN GWs {\color{red} when the window size was not sufficient to resolve them.} Recent developments in the time-frequency analysis (TFA) have yielded various time-frequency representation (TFR) alternatives to the spectrogram. In particular, quadratic TFRs, such as the Wigner--Ville distribution and its modified forms, enable us to perform high-resolution TFA {\color{red} even though those are not free from the trade-off relation of the window selection practically} \citep[e.g.,][]{cohen1995time,stankovic2013time,Boashash2015}.

In this work, we reanalyze a GW prediction from 3D, general-relativistic (GR) core-collapse 
supernova model of a $15M_\odot$ star in \citet{2016ApJ...829L..14K} using both the quadratic TFRs and a (conventional) spectrogram analysis with an adequate window selection. 
We show that the quasistatic low-frequency excess in the spectrogram ($100\sim 300$ Hz) in \citet{2016ApJ...829L..14K}
is clearly decomposed into the two modes, where a dominant mode is at $\sim$ 130 Hz and another is an overtone at $\sim$ 260 Hz with a $\sim$ 10 times smaller amplitude than that of the dominant mode. A new component that was overlooked in \citet{2016ApJ...829L..14K} is also identified whose frequency shows a rather high value of $\sim700$ Hz just after the bounce.
  By employing the same 3D-GR model, \citet{2018MNRAS.477L..96H} recently pointed out that the detection 
  of the GW circular polarization could provide a new probe for pre-explosion hydrodynamics,
 such as the SASI activity and $g$-mode oscillation of the PNS. Going a step further, we develop the quadratic TFA for the GW circular polarization and discuss how significantly we can improve 
  the analysis of the GW polarization compared to that in \citet{2018MNRAS.477L..96H}.
  
This paper is organized as follows. Section \ref{sec:tfa} provides a short summary of 
the waveform employed in this work, followed by a concise overview of the quadratic TFA and
 its application for the waveform analysis. In Section \ref{sec:stokes}, we present an analysis of the GW circular polarization based on the quadratic TFRs. In Section \ref{sec:modes}, we discuss how the GW signatures of the waveform and polarization are related 
 to the emission mechanisms. In Section \ref{sec:noise}, we test the quadratic TFRs using the waveform with detector noises. We summarize the results and discuss its implications in Section \ref{sec:summary}. 

\section{Linear and Quadratic Time-Frequency Analysis of Simulated Gravitational Waves}\label{sec:tfa}

\begin{figure*}[htbp]
\begin{center}
  \includegraphics[bb=0 0 727 172,width=1.0\linewidth]{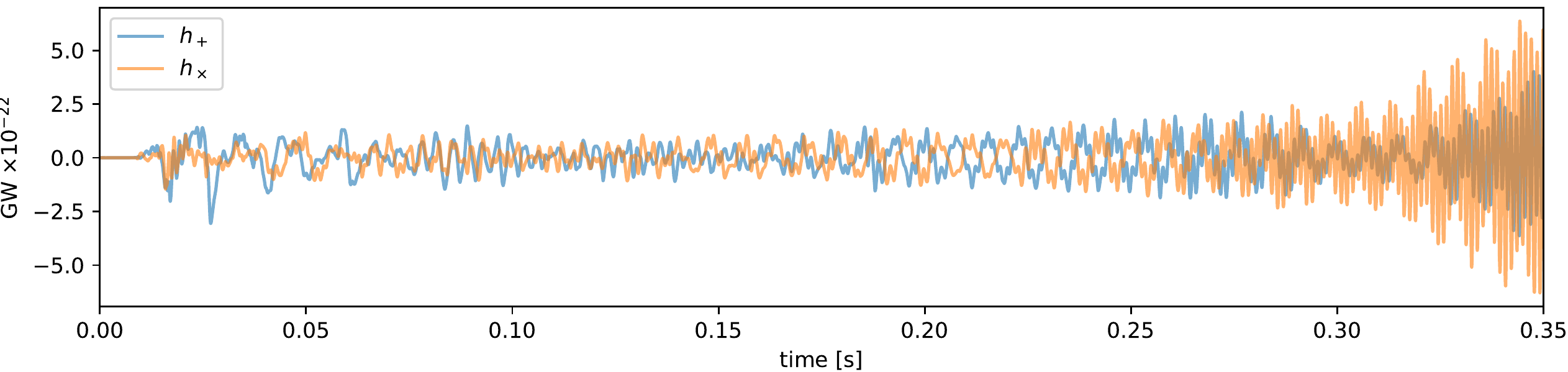}
  \caption{Gravitational waveforms of the $+$ (blue line) and $\times$ (orange line) mode from 
  3D-GR CCSN models of a $15 M_{\odot} $ star (model SFHx in \citet{2016ApJ...829L..14K}). 
   The time is measured after the core bounce (at $t=0$). A source distance of $D = 10$ kpc is assumed.
    \label{fig:yboth}}
\end{center}
\end{figure*}

 In this section, we analyze the GW prediction of model SFHx (Figure \ref{fig:yboth}) from the 3D-GR supernova models in \citet{2016ApJ...829L..14K}.  For this model, the hydrodynamic evolution is self-consistently followed from the onset of core-collapse of a $15M_\odot$ star \citep{WW95}, through to the core bounce, and up to $\sim$ 
350 ms after the bounce. As consistent with the outcomes of recent 3D models (e.g., \citet{Hanke13,Andresen17,Yakunin17}), the 
hydrodynamic evolution is characterized by the prompt convection phase shortly after the bounce
($T_{\rm pb}\lesssim 20$ ms with $T_{\rm pb}$ the postbounce time),
 the linear (or quiescent) phase ($20 \lesssim T_{\rm pb}\lesssim 140$ ms),
 followed by the non-linear phase when the vigorous SASI activity is 
  observed. The dominance of the SASI over neutrino-driven convection 
   persists over $140 \lesssim T_{\rm pb} \lesssim 300$ ms, which is reversed thereafter 
   (see \cite{2016ApJ...829L..14K} for more details).
In the simulation, the BSSN formalism is employed to evolve the metric 
\citep{Shibata95,Baumgarte99}, and the GR neutrino transport is solved 
by an energy-integrated M1 scheme \citep{KurodaT12}. The waveform is extracted 
via a standard quadrupole formula \citet{KurodaT14} where the GR corrections are taken into account.  Since the waveform features are weakly dependent
 on the source orientation (e.g., Figure 4 in \citet{KurodaT17}), 
we focus on the waveforms along an unbiased direction that we choose as
 the north pole ($(\theta,\phi)=(0,0)$) throughout this work.

Regarding the TFA in this paper, all analyses are performed using the Julia package {\it juwvid} (v0.6) \footnote{\color{red}DOI:10.5281/zenodo.1420003}, which was originally developed for axial-tilt measurements of Earth-like exoplanets \citep{2016ApJ...822..112K}. 

\subsection{Time-Frequency Representation (TFR)}\label{ss:if}

The TFR is a function that represents the spectral intensity in the time-frequency domain. TFRs are classified into two major categories. The first contains linear class TFRs, consisting of a Fourier transform of a linear combination of the signal. The second contains quadratic class TFRs that utilize a Fourier transform of the quadratic component of the signal, explained in detail further below. The windowed Fourier transform\footnote{The windowed Fourier transform is also referred to the short-time Fourier transform.} is a representative linear-class TFR defined by 
\begin{eqnarray}
\label{eq:stftdef}
s(\fif, t) = \int_{-\infty}^\infty y(\tau) g(t - \tau,l) e^{- 2 \pi i \fif \tau} d\tau,
\end{eqnarray}
where $y(t)$ is a GW signal and $g(t,l)$ is a window of width $l$. 

\begin{figure*}[htbp]
\begin{center}
  \includegraphics[bb=0 0 383 256,width=0.49\linewidth]{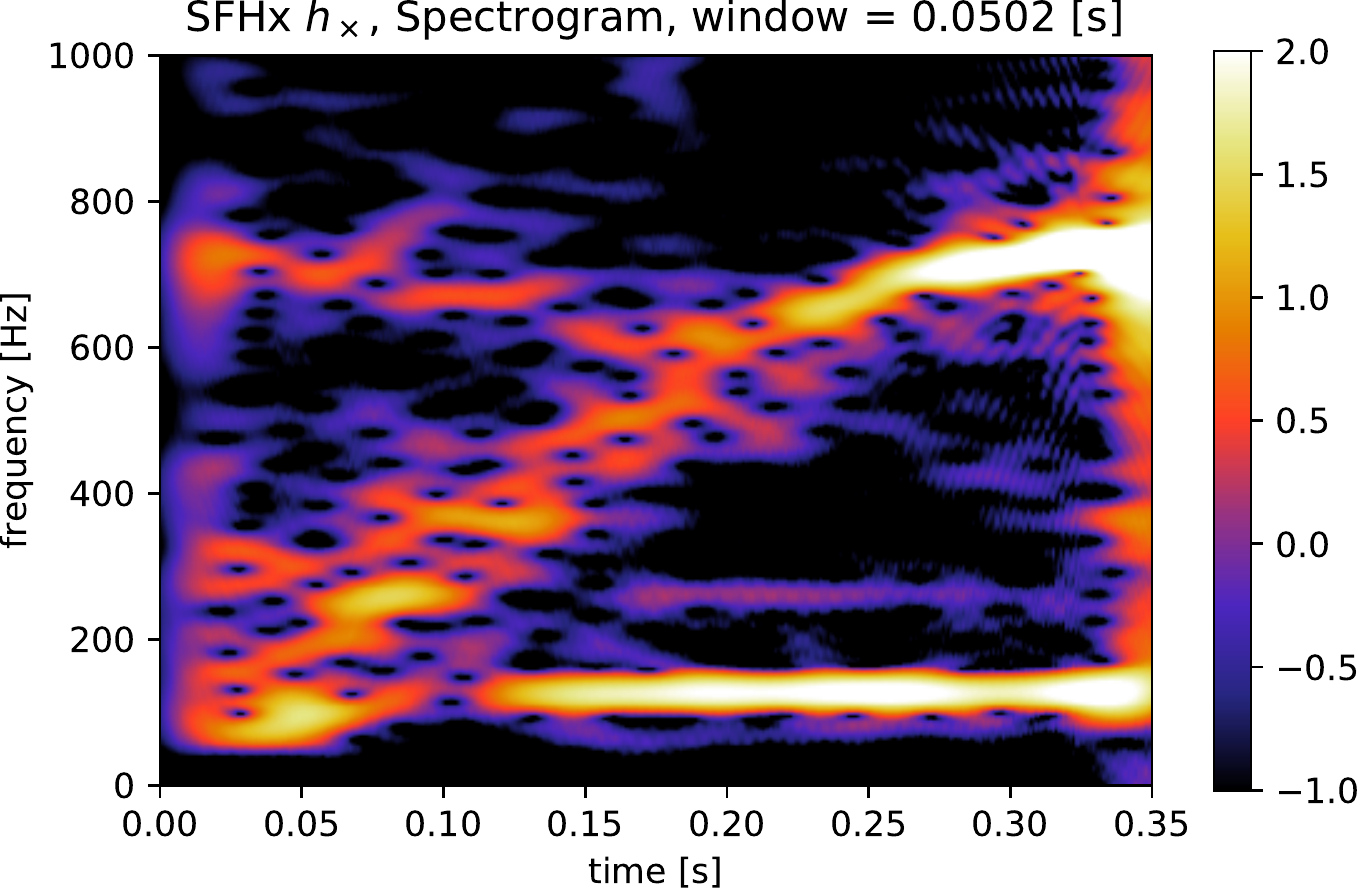}
  \includegraphics[bb=0 0 383 256,width=0.49\linewidth]{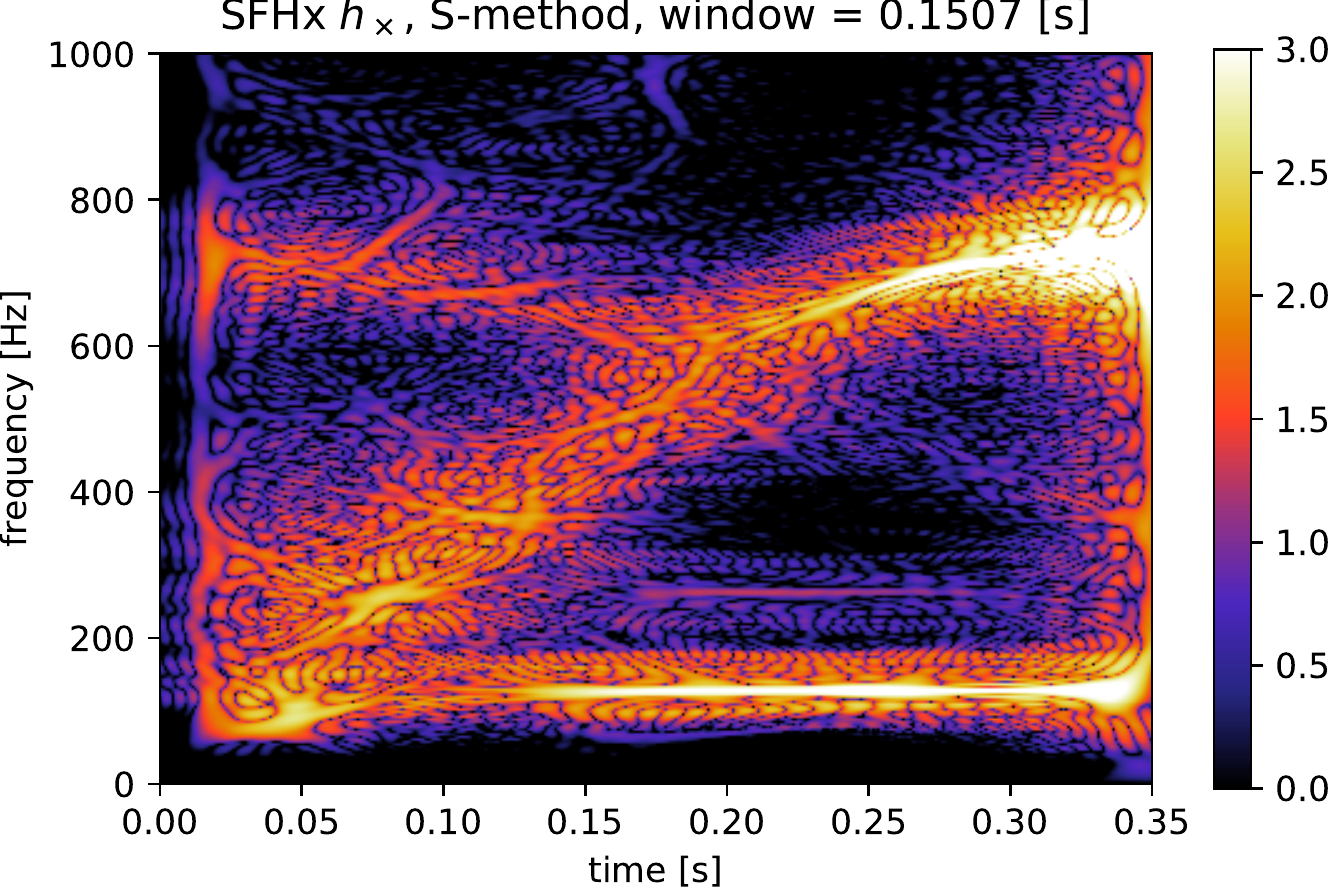}
  \caption{{\color{red} Spectrogram (left) and S-method result (right) for Figure \ref{fig:yboth} ($h_\times$) in a logarithmic color scale. We adopt a window width in equation (\ref{eq:hamming}) of $l = 0.0502$ s for the spectrogram and $l = 0.1507$ s for the S-method and frequency width of  $\theta_L= 24.9$ [Hz] for the S-method. We here use the smaller window for the spectrogram because $l = 0.1502$ s is too large to secure a sufficient time resolution. \label{fig:spectS}}}

\end{center}
\end{figure*}

A window of which the Fourier transform has small side lobes is required to avoid the aliasing effect in the TFR (see Appendix A for the details). In this paper, we use the Hamming window;
\begin{eqnarray}
  \label{eq:hamming}
g(\tau,l) &=& 0.54 + 0.46 \cos{\left(2 \pi \frac{\tau}{l} \right)} \mbox{\,\,\,\,\, for $|\tau| \le l/2$} \nonumber \\
 &=& 0 \mbox{\,\,\,\,\, otherwise.} 
\end{eqnarray}
A spectrogram is expressed as the power of the windowed Fourier transform,
\begin{eqnarray}
\label{eq:spectrogram}
S(\fif, t) \equiv |s(\fif, t)|^2.
\end{eqnarray}
A well-known drawback of the spectrogram is the trade-off relationship between the frequency and time resolution. If a very large window size $l$ is used, the time resolution is less than $\sim l$. In contrast, a small width results in a poor frequency resolution. By changing the window size, we first identify the modes in the spectrogram of the simulated GWs. The left panel in Figure \ref{fig:spectS} shows the corresponding spectrogram. The zero point, $t=0$, corresponds to the time of the core bounce. The main features of the SN GWs in the frequency domain are the multiplicity and nonlinear time evolution of modes. 

The spectrogram is free of artifacts due to the multiplicity of modes, as explained below. However, it is not suitable for the analysis of the fine mode structure or extraction of the instantaneous frequency (IF) because of the low frequency resolution. 

\begin{figure}[htbp]
\begin{center}
  \includegraphics[bb=0 0 406 166,width=\linewidth]{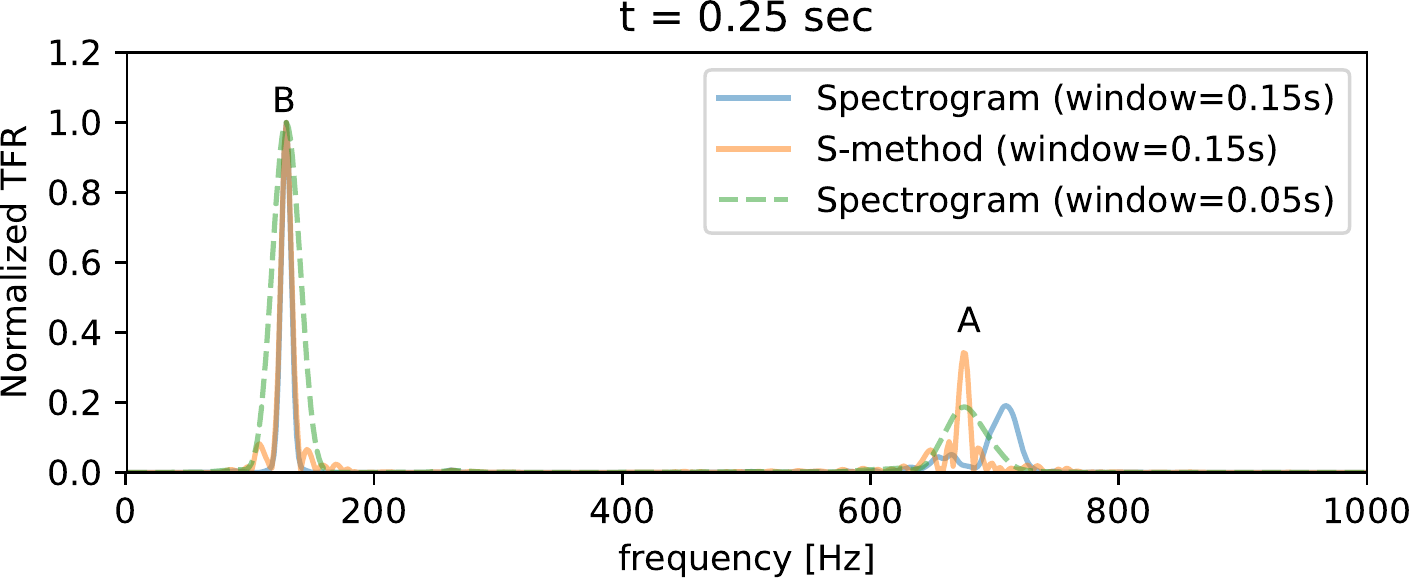}
  \caption{{\color{red}Slice of the TFRs at $t=0.25$s. The orange solid and green dashed lines indicate the slices of the spectrogram (window size= 0.05 s) and the S-method (window size=0.15 s) in Figure \ref{fig:spectS}, respectively. The blue solid line corresponds to the spectrogram with the window size of 0.15 s.\label{fig:slice}}}
\end{center}
\end{figure}

TFRs in the quadratic class are generally suitable for the high-resolution analysis of frequency modes, in particular, the derivation of the IF \citep{cohen1995time}. Let us consider an idealistic case assuming a periodic signal expressed as $z(t) = e^{i \psi(t)}$. The IF is defined as the time derivative of the instantaneous phase of the signal, $\psi (t)$:
\begin{eqnarray}
\mathcal{F} (t) = \frac{1}{2 \pi} \frac{\partial \psi (t)}{\partial t}.
\end{eqnarray}
The idealistic TFR of this signal should be written as
\begin{eqnarray}
\rho (\fif, t) \propto \delta_D (\fif - \mathcal{F}(t)),
\end{eqnarray}
where $\delta_D $ is the Dirac delta function. Then, the inverse Fourier transform of $\rho (\fif, t)$ in terms of $\fif$ and $\tau$ is given by
\begin{eqnarray}
\label{eq:hatrho}
\hat{\rho} (\tau, t) = e^{2 \pi i \mathcal{F} (t) \tau} = \exp{\left( i \tau \frac{\partial \psi (t)}{\partial t}\right)}.
\end{eqnarray}
If we adopt the linear approximation of the IF in terms of time then; 
\begin{eqnarray}
\label{eq:lfmwv}
 \frac{\partial \psi (t)}{\partial t} \approx  \frac{\psi (t + \tau/2) - \psi (t - \tau/2)}{\tau},
\end{eqnarray}
and we obtain the Wigner--Ville distribution, 
\begin{eqnarray}
\rho (\fif, t) &=& \int_{-\infty}^\infty \hat{\rho} (\fif, \tau) e^{- 2 \pi i \fif \tau} d \tau \\
&\approx& \int_{-\infty}^\infty e^{i \psi (t + \tau/2) - i \psi (t - \tau/2)} e^{- 2 \pi i \fif \tau} d \tau \\
\label{eq:wd}
&=& \int_{-\infty}^\infty z(t+\tau/2) z^\ast(t-\tau/2) e^{- 2 \pi i \fif \tau} d \tau \nonumber \\
&\equiv& \rho_\mathrm{wv} (\fif, t),
\end{eqnarray}
which is the most prominent member of the quadratic class. For a real-valued signal, $y(t) \in \mathbb{R}$, the commonly used analytic signal, $z(t) \in \mathbb{C}$, is defined as
\begin{eqnarray}
\label{eq:zt}
z(t) = \frac{2}{2 \pi} \int_0^\infty d \omega \tilde{y}(\omega) e^{i \omega t},
\end{eqnarray}
where $\tilde{y} (\omega)$ is the Fourier transform of $y$ \citep[e.g.,][]{cohen1995time,stankovic2013time,Boashash2015}.

As the nonlinearity of the IF induces artifacts in the Wigner--Ville distribution \citep{cohen1995time}, the pseudo Wigner--Ville (PWV) distribution is used in practice: 
\begin{eqnarray}
  \label{eq:pwv}
\rho_\mathrm{pwv} (\fif, t) = \int_{-\infty}^\infty w(\tau) z(t+\tau/2) z^\ast(t-\tau/2) e^{- 2 \pi i \fif \tau} d \tau, \nonumber 
\end{eqnarray}
where $w(\tau)$ is the window. The window width should be smaller than the typical scale of the IF fluctuation of interest, obtained from the linear trend \citep[see Appendix D in ][]{2016ApJ...822..112K}.  However, the PWV distribution of multimode signals exhibits significant artifacts owing to the presence of a non-zero cross term between multiple modes \citep{stankovic2013time}. The reason for the appearance of the non-zero cross term in the PWV distribution is as follows. Assuming a symmetric window, $w(\tau) = g(-\tau/2) g^\ast(\tau/2)$, we obtain  
\begin{eqnarray}
&\rho_\mathrm{pwv}& (\fif, t) \nonumber \\
&=& 2 \int_{-\infty}^\infty g(-\tau^\prime) g^\ast(\tau^\prime) z(t+\tau^\prime) z^\ast(t-\tau^\prime) e^{- 2 \pi i (2 \fif) \tau^\prime} d \tau^\prime \nonumber \\
&=&  4 s (2 \fif, t) \ast s^\ast (2 \fif, t) \\
  \label{eq:naivesm}
&=& 2 \int_{-\infty}^\infty s (\fif+\theta/2, t) s^\ast (\fif-\theta/2, t) d\theta,
\end{eqnarray}
where $\tau^\prime = \tau/2$. Suppose that the signal has two modes, $\fif_1$ and $\fif_2$. Then, $ s (\fif_1, t) s^\ast (\fif_2, t) $ in the integral of equation (\ref{eq:naivesm}), and becomes a non-zero term because both $s(\fif_1,t)$ and $s(\fif_2,t)$ are significant values. Solving $\fif_1 = \fif_c+\theta/2$ and $\fif_2 = \fif_c - \theta/2$, we find that the non-zero cross term between the two signals appears at $\fif_c = (\fif_1 + \fif_2)/2$. 

\begin{figure}[htbp]
\begin{center}
  \includegraphics[bb=0 0 386 276,width=\linewidth]{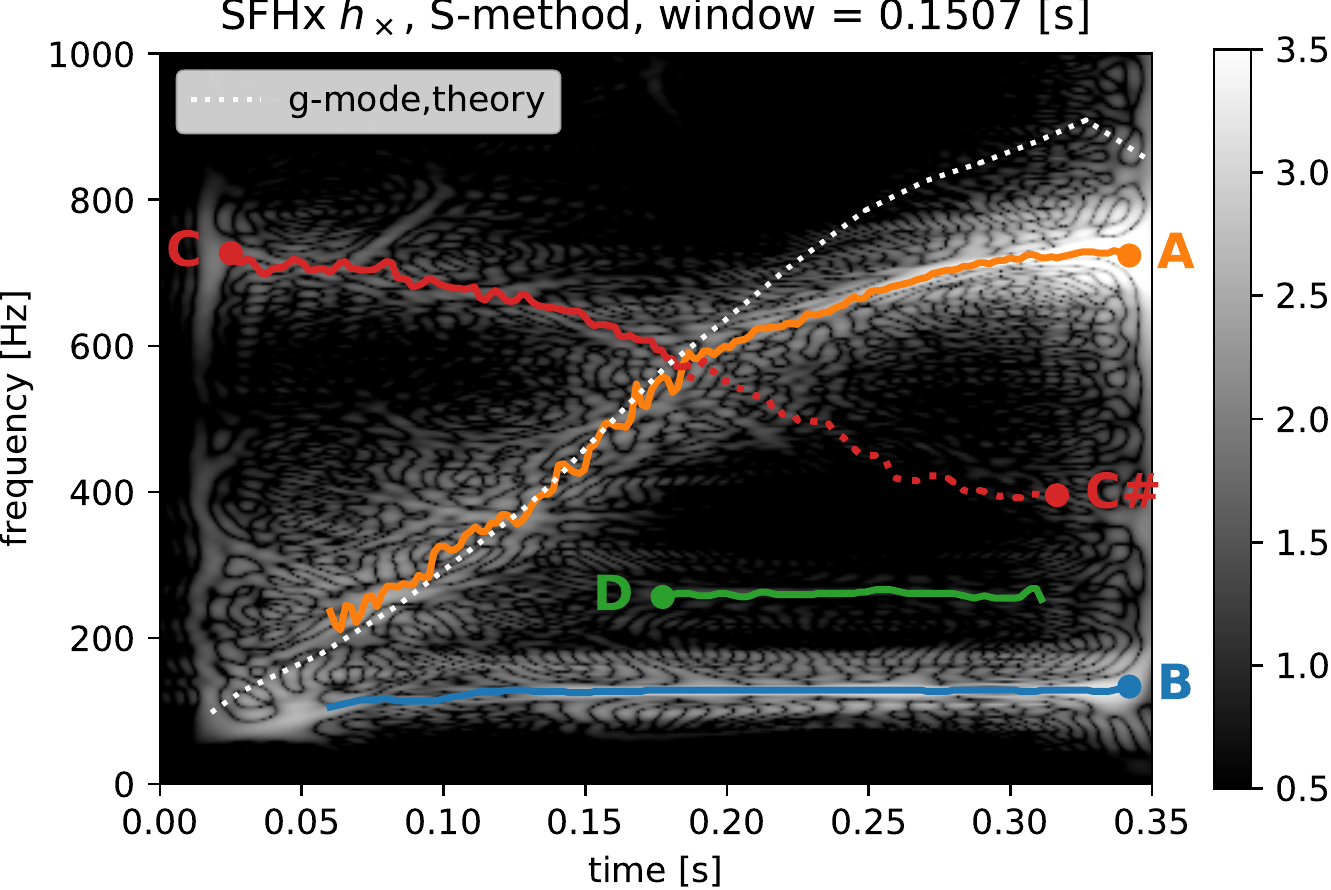}
  \caption{Mode identification by IF tracking via the S-method. {\color{red} The theoretical prediction of the PNS $g$-mode oscillation, which represents the peak frequency according to \citet{BMuller13}, is shown by the white dashed curve.}   \label{fig:smif}}
\end{center}
\end{figure}

To overcome the interference between modes, the S-method was developed by \citet{1994ITSP...42..225S}. The S-method introduces a new window $P(\theta)$ in the frequency domain in equation (\ref{eq:naivesm}) as
\begin{eqnarray}
\label{eq:smeth}
\rho_\mathrm{s} (\fif, t) &=& 2 \int_{-\infty}^\infty P(\theta) s (\fif+\theta/2, t) s^\ast (\fif-\theta/2, t) d\theta \\
\label{eq:smeth2}
 &=& 2 \int^{\theta_L}_{-\theta_L} s (\fif+\theta/2, t) s^\ast (\fif-\theta/2, t) d\theta,  
\end{eqnarray}
where we adopt a boxcar window for $P(\theta)$, and $\theta_L$ is the width of the frequency window.

The right panels of Figure \ref{fig:spectS} show the TFRs of the S-method. The TFRs in Figure \ref{fig:spectS} exhibit two quasistatic modes around 130 and 260 Hz, an increasing mode, and a decreasing mode. These modes are then identified by tracking the IF.  { \color{red}  To see the difference between those TFRs, we show the slices at $t = 0.2$ sec in Figure \ref{fig:slice}.  Although the S-method does not improve the sharpness of mode B, it does increase the resolution of mode A, and the peak value of the S-method is slightly shifted compared with that of the spectrogram. Although the spectrogram with the longer window of 0.15sec exhibits a similar resolution to that of the S-method, the peak value is slightly shifted. Notably, the peak of the spectrogram using the short window of 0.05sec, i.e., a good time resolution, converges to that of the S-method although the peak has less sharpness. It means that the spectrogram exhibits some artifact of the center of modes for the high-frequency resolution. These features are crucial when we identify each mode as discussed next. }

{\color{red} Table \ref{tab:quad} summarizes the relation between three TFRs, the spectrogram, the PWV, and the S-method.}

\begin{table}[!tbh]
\begin{center}
\caption{\label{tab:quad}}
\begin{tabular}{ccc}
  \hline\hline
  \multicolumn{3}{c}{TFR defined by equation (\ref{eq:smeth2})} \\
  \hline
    $\theta_L \to 0$ & $0 < \theta_L < \infty$ & $\theta_L \to \infty$  \\
  spectrogram & S-method & PWV distribution\\
  eq. (\ref{eq:spectrogram}) &  & eq. (\ref{eq:pwv}) \\ 
  \hline
  \multicolumn{3}{c}{Stokes TFR defined by equation (\ref{eq:smethStokes2})} \\
  \hline
    $\theta_L \to 0$ & $0 < \theta_L < \infty$ & $\theta_L \to \infty$  \\
  spectrogram-type & S-type & PWV-type \\
  eq. (\ref{eq:spectStokes}) &  & eq. (\ref{eq:pwvStokes}) \\
\end{tabular}
\end{center}
\end{table}

\begin{figure}[htbp]
\begin{center}
  \includegraphics[bb=0 0 461 346,width=\linewidth]{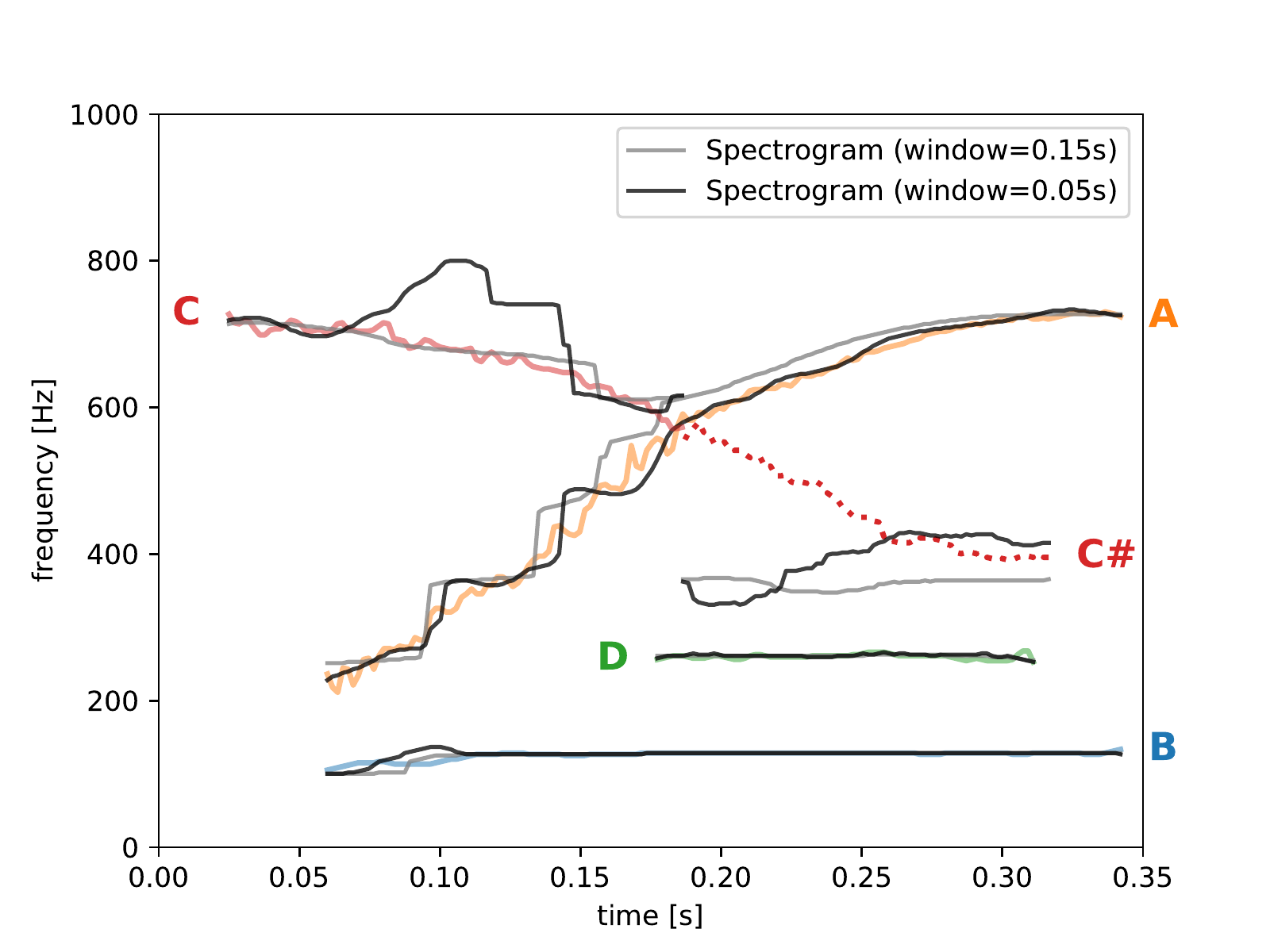}
  \caption{\color{red} Comparison of the mode identification by the spectrograms and the S-method. The gray lines are the IF tracked by the spectrogram with different window sizes; light and dark lines corresponding to the window size of 0.15 sec and 0.05 sec, respectively. The color lines are the IF by the S-method, same as Figure \ref{fig:smif}. \label{fig:smif_diff}}
\end{center}
\end{figure}

\subsection{Mode Identification \label{ss:modeid}}

To identify the modes, we track the IF of the signals by the S-method. The IF is derived from the first moment of the TFR or a ridge line of the TFR \citep[e.g.,][]{Boashash2015}. Because we have multiple modes in the TFR, we need to limit the integration section to a width, $H$, for the computation of the first moment of the TFR at time $t$;
\begin{eqnarray}
  \label{eq:interate}
  \fif_c^{(i+1)} (t) = \frac{\int_{\fif_c^{(i)} (t) - H/2}^{\fif_c^{(i)} (t)+ H/2} \fif \rho_\mathrm{s} (\fif, t) d \fif }{\int_{\fif_c^{(i)} (t) - H/2}^{\fif_c^{(i)} (t) + H/2} \rho_\mathrm{s} (\fif, t) d \fif}.
\end{eqnarray}
Equation (\ref{eq:interate}) is iterated of for an index $i$ until $\fif_c^{(i)} (t)$ converges, yielding the IF $\fif_c^{\circ} (t_j)$ at $t_j$. We identify the starting point of $\fif_c^{(0)} (t)$ around a clear peak of TFR at time $t=t_0$ and derive $\fif_c^{\circ} (t_0)$ by iteration of equation (\ref{eq:interate}). We use $\fif_c^{\circ} (t_j)$ as the initial guess for the next time step, $t_{j+1}$:
\begin{eqnarray}
 \fif_c^{(0)} (t_{j+1}) = \fif_c^{\circ} (t_j).
\end{eqnarray}
Because this procedure tracks the IF of a mode in the TFR, we refer to it as {\it IF tracking}. Figure \ref{fig:smif} shows the results of IF tracking by the S-method. The initial guess $\fif_c^{\circ} (t_0)$ for each mode is shown by a filled circle. We clearly detected five modes, A B, C, $\mathrm{C}^\#$, and D. 

{\color{red} As the agreement with the theoretical prediction in Figure \ref{fig:smif} suggests}, mode A corresponds to the g-mode oscillation, which is excited at the PNS surface by the downflows and/or directly originates from the deceleration of infalling convective plums \citep{BMuller13}. Mode C intersects mode A at $t \sim 0.18$ s. Mode C is not tracked after the intersection because mode A is much stronger than mode C. Hence, we start to track mode C again from the opposite side ($\mathrm{C}^\#$). Although visual inspection suggests that the mode $\mathrm{C}^\#$ is not significant in either the spectrogram or S-method results, the IF tracking clearly shows that this mode is smoothly connected to mode C. Mode $\mathrm{C}^\#$ is likely to be an extension of mode C. Note in \citet{2016ApJ...829L..14K}, the window of the spectrogram was too small to separate modes B and D owing to the trade-off relationship between the frequency and time resolution. 

{\color{red} In Figure \ref{fig:smif_diff}, we also compare the mode identification by the spectrogram (window widths of 0.05 and 0.15 sec) with that by the S-method (the colored lines). The both IFs are similar for the constant modes, i.e., for modes B and D. However, the IFs of modes A and C by the spectrogram exhibit stepped patterns due to a poor time resolution. The spectrogram with a smaller window of 0.05 sec (the dark gray lines) slightly reduces this artifact. However, the decrease of the frequency resolution makes challenging to track the IF as appeared in mode C. Also, the spectrograms are not able to track the weaker mode C\#.
}

\section{Quadratic Time-Frequency Representations of Polarization States}\label{sec:stokes}

{\color{red} In the previous section, we applied the standard methodology of the linear and quadratic time-frequency analysis \citep[e.g.][]{cohen1995time,stankovic2013time,Boashash2015} to the simulated GWs. In this section, we develop the TFR of polarization states. In addition to the standard polarization analysis that uses the short-time Fourier analysis, we introduce the quadratic version of the polarization analysis as original research.} Because the polarization is connected to the mutual phase shift between $h_+$ and $h_\times$, these need to be correlated. The Stokes parameters are normally defined as  
\begin{eqnarray}
I (\fif, t) &=& W_{11} + W_{22}, \\
Q (\fif, t) &=& W_{11} - W_{22}, \\
U (\fif, t) &=& W_{12} + W_{21}, \\
V (\fif, t) &=& i (W_{21} - W_{12}),
\end{eqnarray}
where $W_{ij}$ is the cross term of two signals. The short-time Fourier transform is conventionally used to define the cross term \citep[e.g.,][]{2007PhRvL..99l1101S,2016PhRvL.116o1102H,2018MNRAS.477L..96H},
\begin{eqnarray}
\label{eq:spectStokes}
  W_{ij} (\fif, t) = s_i(\fif, t) s_j^\ast(\fif, t),
\end{eqnarray}
where $s_i(\fif, t)$ and $s_j(\fif, t)$ are the short-time Fourier transforms of the real signals $h_+ (t)$ and $h_\times (t)$, respectively. In this paper, we refer to the Stokes parameters obtained using the cross term of equation (\ref{eq:spectStokes}) as spectrogram-type Stokes parameters.

Here, we define the quadratic form of the Stokes parameters by extending the S-method to the cross form of the TFR;
\begin{eqnarray}
\label{eq:smethStokes}
W_{ij} &\equiv& 2 \int_{-\infty}^\infty P(\theta) s_i (\fif+\theta/2, t) s_j^\ast (\fif-\theta/2, t) d\theta \\
\label{eq:smethStokes2}
&=& 2 \int_{-\theta_L}^{\theta_L} s_i (\fif+\theta/2, t) s_j^\ast (\fif-\theta/2, t) d\theta.
\end{eqnarray}
We refer to the Stokes parameters from equation (\ref{eq:smethStokes}) as the S-type Stokes parameters.

The limit $\theta_L \to 0$ or $P(\theta) \to \delta_D(\theta)$ of equation (\ref{eq:smethStokes}) yields equation (\ref{eq:spectStokes}). In the limit $\theta_L \to \infty$, equation (\ref{eq:smethStokes}) converges to the cross PWV distribution,
\begin{eqnarray}
  \label{eq:pwvStokes}
W_{ij} = \int_{-\infty}^\infty w(\tau) z_i(t+\tau/2) z_j^\ast(t-\tau/2) e^{- 2 \pi i \fif \tau} d \tau, 
\end{eqnarray}
where $z_i$ and $z_j$ are the analytic signals of $h_+ (t)$ and $h_\times (t)$, respectively. We refer to the Stokes parameters obtained using equation (\ref{eq:pwvStokes}) as PWV-type Stokes parameters. 

We show that the PWV-type Stokes parameters have the same properties as the conventional Stokes parameters used in electromagnetism. {\color{red} Suppose that the analytic signals $z_1=A_1 e^{i \psi_1(t)}$ and $z_2=A_2 e^{i \psi_2(t)}$ } both have a single mode with the IF
\begin{eqnarray}
\mathcal{F}_{1} (t) &=& \frac{1}{2 \pi} \frac{\partial \psi_1 (t)}{\partial t},\\
\mathcal{F}_{2} (t) &=& \frac{1}{2 \pi} \frac{\partial \psi_2 (t)}{\partial t},
\end{eqnarray}
respectively. First, we can express the intensity as
\begin{eqnarray}
  I (\fif, t) &\approx& A_1^2 \delta_D (\fif - \mathcal{F}_{1}(t)) +  A_2^2 \delta_D (\fif - \mathcal{F}_{2}(t)) \\
  &\approx& (A_1^2+A_2^2) \delta_D (\fif - \mathcal{F}(t)).
\end{eqnarray}
In the last equation, we assume that the IFs of the two modes are similar,
\begin{eqnarray}
\mathcal{F} (t) \equiv \mathcal{F}_{1} (t) \approx \mathcal{F}_{2} (t).
\end{eqnarray}
In this limit, the Stokes $I$ parameter represents the summation of the square norm of the signals over the IF of $\mathcal{F} (t)$. Likewise, we obtain
\begin{eqnarray}  
  Q (\fif, t) &\approx& A_1^2 \delta_D (\fif - \mathcal{F}_{1}(t)) - A_2^2 \delta_D (\fif - \mathcal{F}_{2}(t)) \\
  &\approx& (A_1^2 - A_2^2) \delta_D (\fif - \mathcal{F}(t)),
\end{eqnarray}
which means that $Q$ represents a $0^\circ$ ($90^\circ$) linear polarization. In addition, 
\begin{eqnarray}  
&U& (\fif, t) 
  = \int_{-\infty}^\infty A_1 A_2 e^{i \psi_1 (t + \tau/2) - i \psi_1 (t - \tau/2)} \nonumber \\
  &\times& \left[ e^{- i \Delta \psi(t - \tau/2) } + e^{ i \Delta \psi(t+\tau/2)}  \right] e^{- 2 \pi i \fif \tau} d \tau \nonumber \\
  &\approx& \int_{-\infty}^\infty 2 A_1 A_2 (\cos{\Delta \psi}) [e^{i \psi_1 (t + \tau/2) - i \psi_1 (t - \tau/2)}] e^{- 2 \pi i \fif \tau} d \tau \nonumber \\
  &\approx& \int_{-\infty}^\infty 2 A_1 A_2 (\cos{\Delta \psi}) e^{2 \pi i \mathcal{F} (t) \tau} e^{- 2 \pi i \fif \tau} d \tau \nonumber \\
  &\approx& 2 A_1 A_2 (\cos{\Delta \psi}) \delta_D (\fif - \mathcal{F}(t))   
\end{eqnarray}
represents a $45^\circ$ ($135^\circ$) linear polarization, {\color{red} where $\Delta \psi \equiv \psi_2 - \psi_1$}. A similar derivation shows that the $V$ parameter represents a circular polarization,
\begin{eqnarray}  
  V (\fif, t) &\approx& 2 A_1 A_2 (\sin{\Delta \psi}) \delta_D (\fif - \mathit{f}_{i}(t)).
\end{eqnarray}

\begin{figure*}[htbp]
\begin{center}
  \includegraphics[bb=0 0 611 407,width=1.0\linewidth]{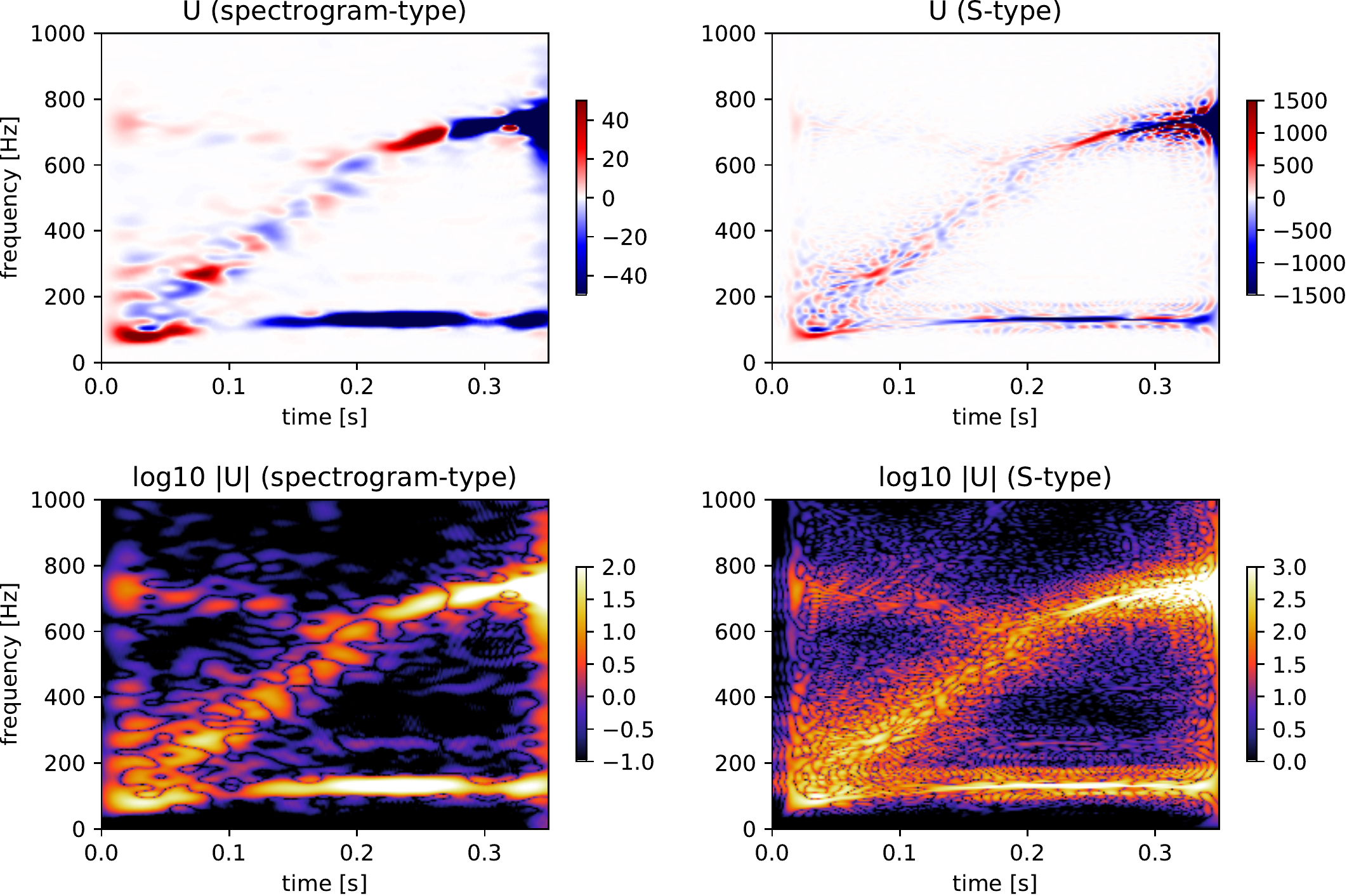}
  \caption{Stokes parameters in TFRs ($U$ mode) of Figure \ref{fig:yboth}. {\color{red} The left and right upper panels show the spectrogram-type (window size of 0.15sec)  and S-type (window size of 0.05sec) Stokes $U$ parameters}, respectively. The bottom panels show {\color{red}$|U|$} from the upper panels on a logarithmic scale. The time and frequency windows are the same as those in Figure \ref{fig:spectS}. \label{fig:compV4}}
\end{center}
\end{figure*}

Figure \ref{fig:compV4} shows the difference between the Stokes U parameter of the spectrogram type (left) and the S type (right) with the same time window width (0.15 s). {\color{red} Although the polarization modes obtained using the S-type Stokes parameters exhibit a higher resolution than those obtained using the spectrogram-type ones, there is the grainy structure in the S type TFR. Therefore, we choose the spectrogram-type Stokes parameters to see the characteristics of the GW polarization.}


\begin{figure*}[htbp]
\begin{center}
  \includegraphics[bb=0 0 599 407,width=1.0\linewidth]{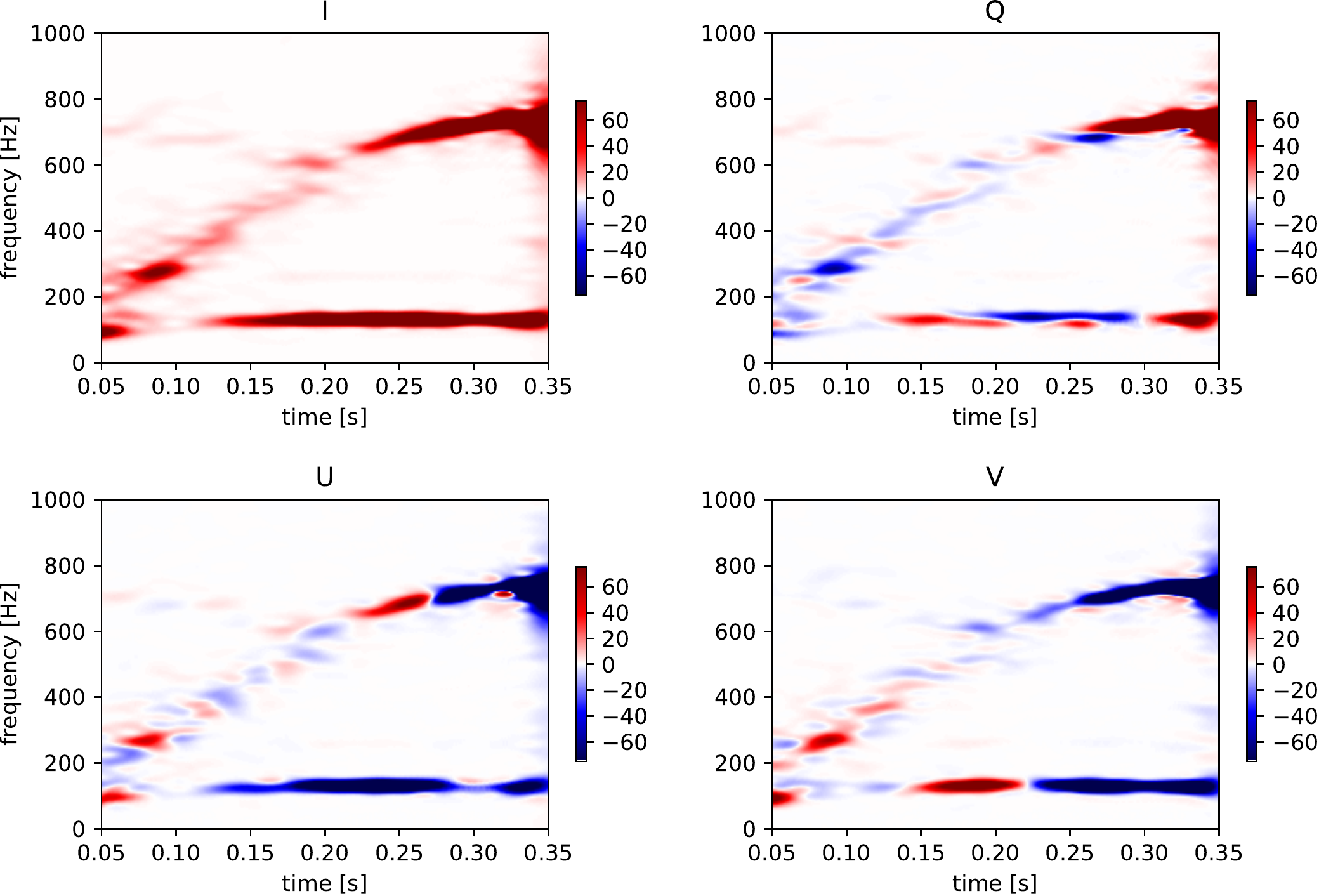}
  \caption{The spectrogram-type Stokes parameters; $I, Q, U,$ and $V$, in order from upper left to bottom right.  \label{fig:allstokes}}
\end{center}
\end{figure*}

As shown in Figure \ref{fig:allstokes}, the polarization states of mode B change slowly on a time scale on the order of 0.1 s, whereas mode A exhibits more rapid random changes. We suggest that these features reflect the physical origins of the modes, which we will discuss in the next section in more detail.

\section{Nature of Modes A--D}\label{sec:modes}

\begin{figure}[htbp]
\begin{center}
  \includegraphics[bb=0 0 414 593,width=\linewidth]{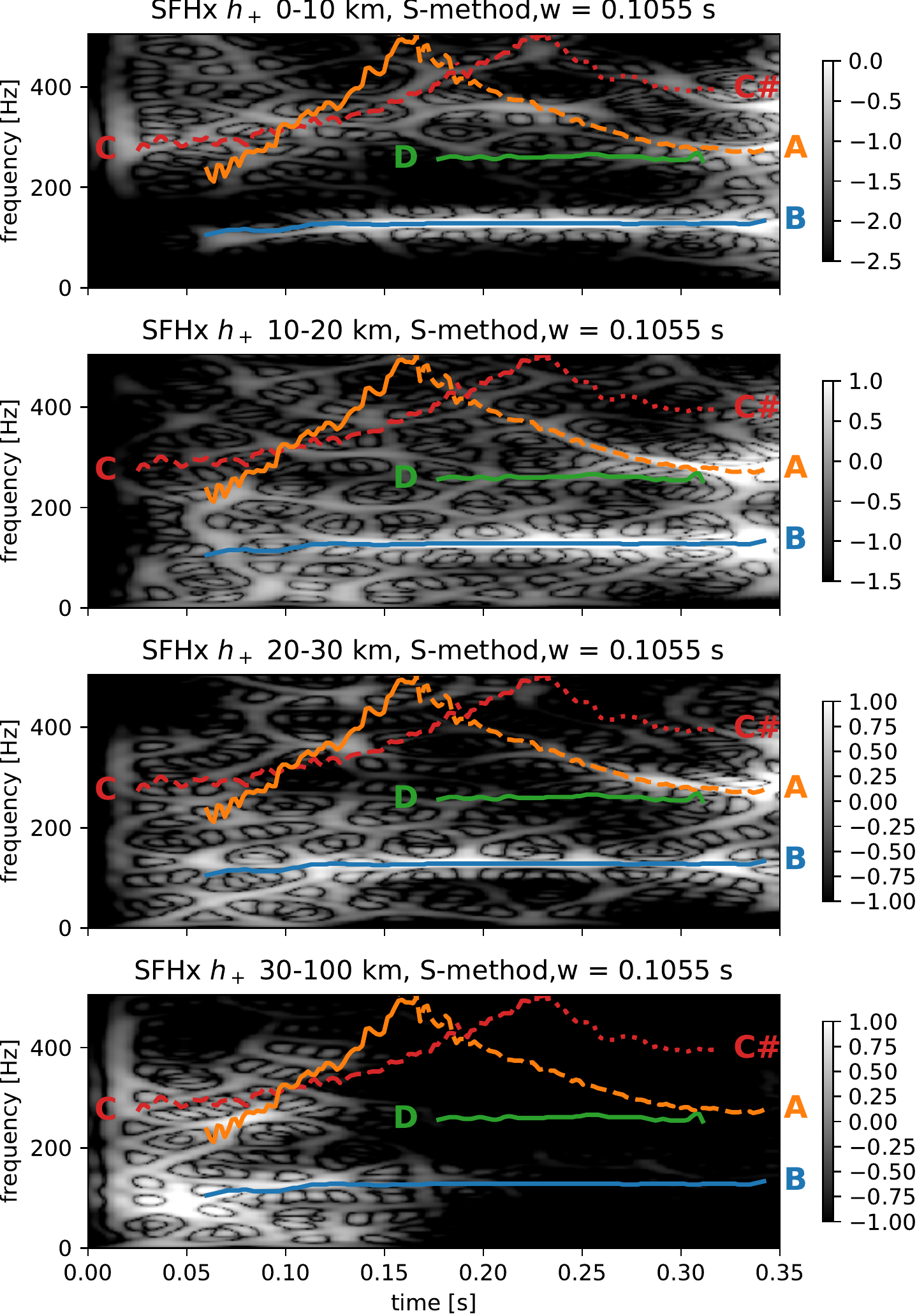}
  \caption{Right: S-method result of the GWs from each shell. Left: Extracted IFs overlaid onto the right panels. The colors are the same as in Figure \ref{fig:smif}. The dashed lines are the aliased modes of A and C for $n=1$. {\color{red} The Nyquist frequency of the simulations is $f_{\mathrm{Ny}} = 500$ Hz.} \label{fig:shell}}
\end{center}
\end{figure}

In this section, we discuss modes A--D individually. First, we identify the locality of each mode by analyzing the radial shells used in \citet{2016ApJ...829L..14K} of $0 < r < 10$ km, $10 < r < 20$ km, $10 < r < 30$ km, and $30 < r < 100$ km, as shown in Figure \ref{fig:shell}. Unfortunately, we have only low-cadence shell data with a Nyquist frequency of $f_{\mathrm{Ny}}=500$ Hz (i.e., 1 ms sampling). The modes above the Nyquist frequency appear as a false mode with an aliased frequency $\tilde{f}_n (t)$. The relationship between the true and aliased frequencies is expressed as 
\begin{eqnarray}
f_t (t) = |2 n f_{\mathrm{Ny}} - \tilde{f}_n (t)|,
\end{eqnarray}
where $\tilde{f}_n(t)$ is the aliased frequency for $n=\pm 1, \pm 2,..,$, and $f_{\mathrm{Ny}}$ is the Nyquist frequency (500 Hz). Hence, we can analyze the $n=1$ aliased signature for the high-frequency parts of A and C above 500 Hz as a proxy of the true frequency\footnote{The widespread misunderstanding that information above the Nyquist frequency cannot be obtained is not true. Instead, the true frequency cannot be distinguished from the aliased signatures without knowledge of the modes above the Nyquist frequency. In our case, we have a priori information on modes A and C from the original analysis of the waveform.}. 

Therefore, the strong feature around 350 Hz at $t=0.25$--$0.35$ s can be interpreted as an aliased signature of mode A at 700 Hz ($n=1$, $f_{\mathrm{Ny}}=500$ Hz), as shown by the orange dashed line in Figure \ref{fig:shell}. Modes A--C originate from the PNS ($< 30$ km). Although it is still not clear, {\color{red} mode C is likely to arise in the innermost region ($\sim10$ km) from spatial decomposition analysis. The emission mechanism of mode C might be one of the $p$-modes oscillation as it shows consistent frequency range, evolution, and spatial origin in the results of \citep{torres18b}. We need detailed linear analysis to determine the physical origin that leads to mode C.} Mode D is not clear in some panels owing to contamination of the aliased frequency of mode A.

\subsection{Modes A and C}

The theoretical curve \citep[taken from ][]{2016ApJ...829L..14K} of the PNS $g$-mode oscillation agrees well with mode A until it crosses over mode C ($t \sim 0.2 $ s). The Stokes parameters of both modes A and C before $t \sim 0.2$ s exhibit a random rapid change in the $Q$, $U$, and $V$ modes. This rapidly and randomly changing polarization state
is a natural outcome of the buoyancy-driven convection (in association with the PNS $g$-mode 
 oscillation) at least before $t \sim 0.2$ s. {\color{red} Likewise, mode C is likely to be connected to the convection excitation of the PNS. Although yet to be determined, this mode is likely to arise from one of the $p$-modes oscillation from its frequency range, evolution, and spatial origin \citep{torres18b}.}
Evidently, a more detailed analysis is required to connect mode C with the PNS convection, as the match does not always hold throughout the postbounce phase.
After $t \sim 0.25$ s, the polarization state of mode A becomes quasistatic, as for modes B and D.
The deviation of the IF of mode A from the 
 linear analysis becomes more significant in the 
 later postbounce time (Figure \ref{fig:smif}), which was also observed in \citet{BMuller13,viktoriya18}. 

\subsection{Modes B and D}\label{ss:BD}

\citet{2016ApJ...829L..14K} argued that the quasistatic mode originates from quadratic deformation of a PNS core surface with $\rho\sim10^{14}$ g cm$^{-3}$ due to the mass accretion by the SASI. This is because the violent SASI motion triggers a remarkable time modulation in the mass accretion rate onto the PNS core surface. The time modulation in the mass accretion rate thus basically shows the same time frequency as the SASI. Such a quasistatic mode was also reported in \citet{Andresen17}. As discussed in the previous section, this quasistatic mode consists of two modes, modes B ($f_B=$ 130 Hz) and D ($f_D=$ 260 Hz). {\color{red} The dominant frequency of the SASI in our simulation is $f_\mathrm{SASI} \sim$ 65 Hz \citep{2016ApJ...829L..14K}. The fact that $f_\mathrm{SASI}$ is approximately half of $f_B$ is indicatative of the connection between the mode B and the SASI.}

\begin{figure*}[htbp]
\begin{center}
\includegraphics[bb=0 0 504 360,width=0.51\linewidth]{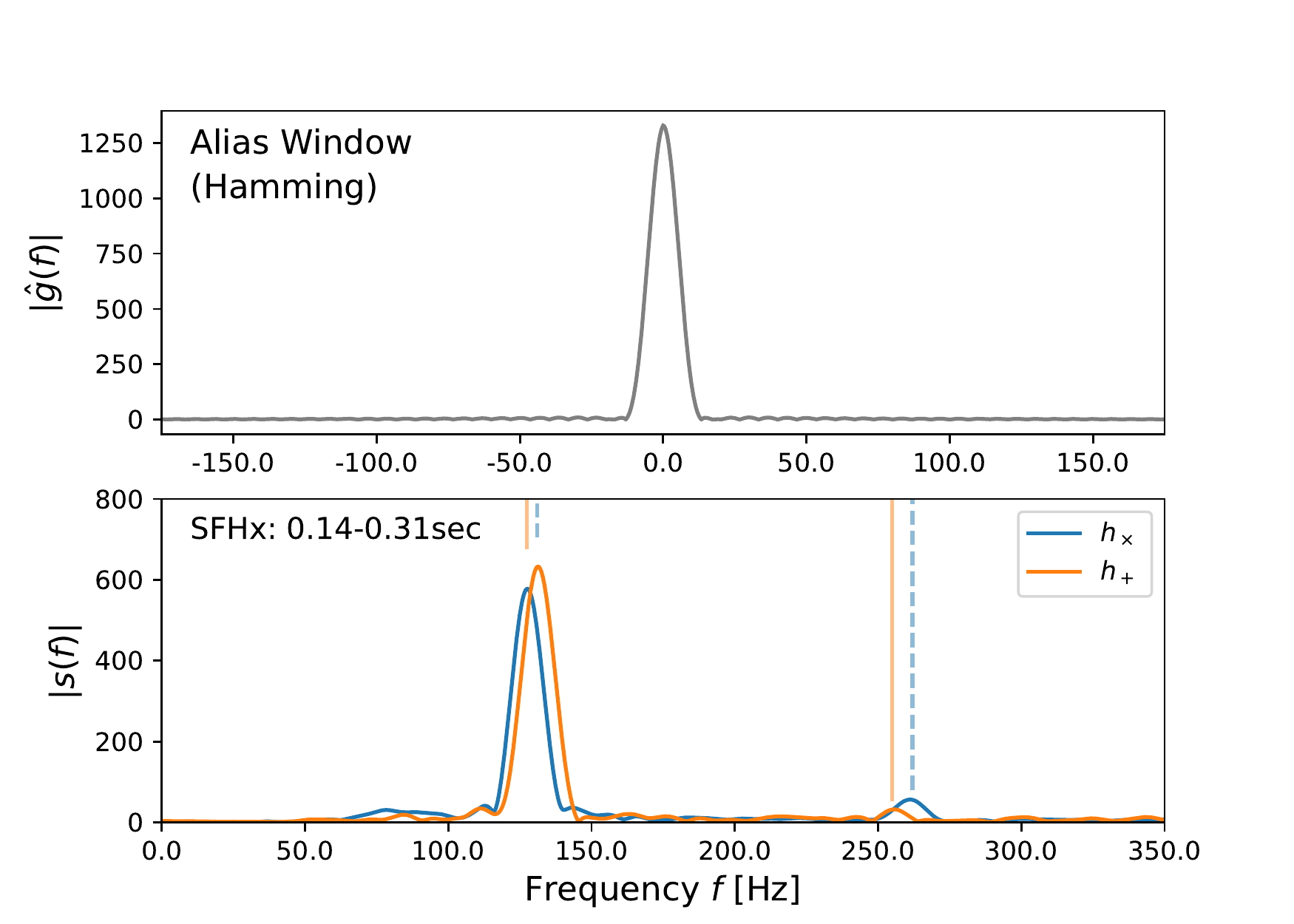}
\includegraphics[bb=0 0 461 346,width=0.48\linewidth]{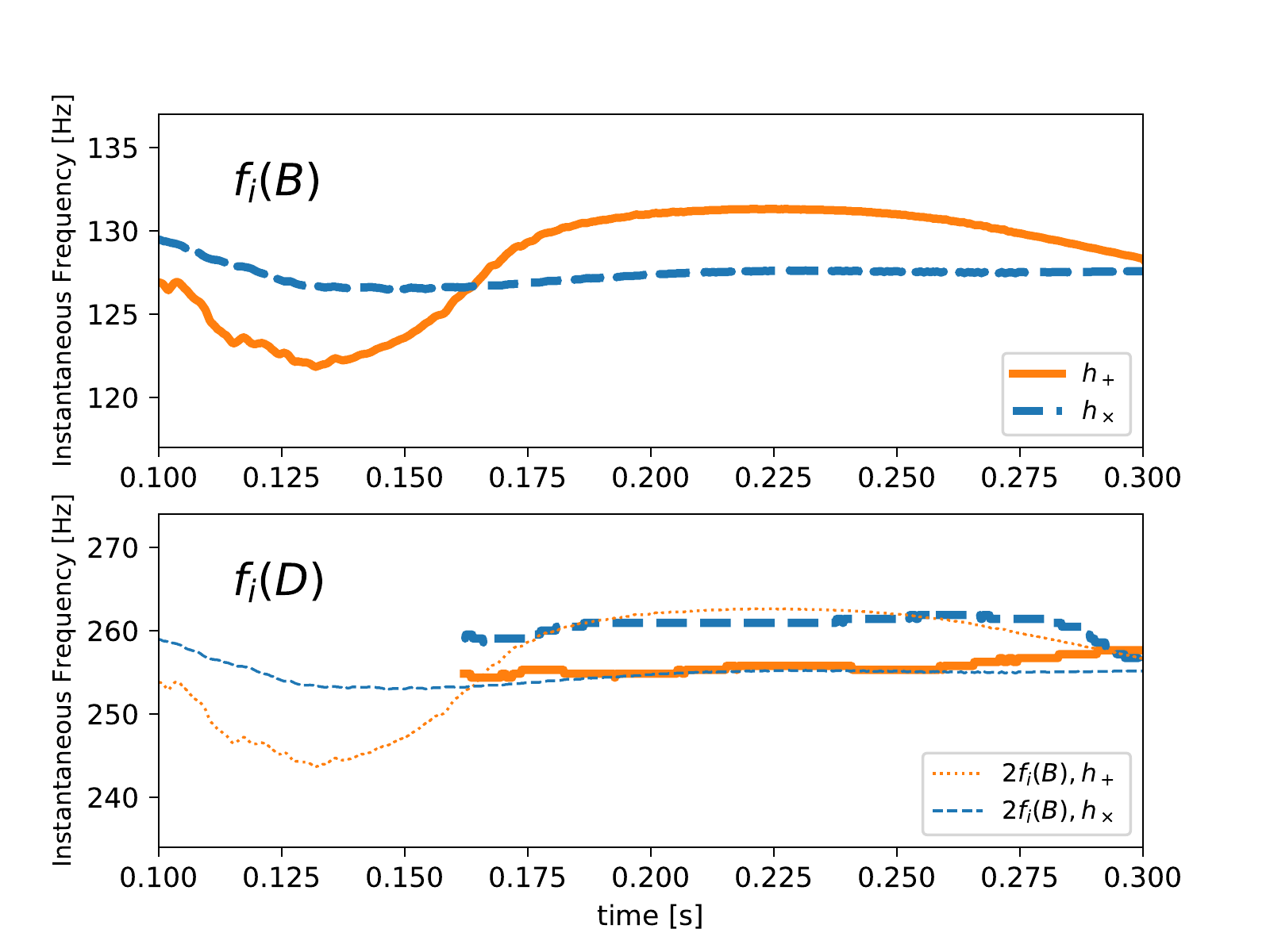}
\caption{Spectra of modes B and D. The top left panel shows an aliased window of the Hamming window, i.e., the Fourier amplitude of the Hamming window $¥hat{g} (f)$. The bottom left panel shows the Fourier amplitude of GWs between 0.14 and 0.31 s. The solid (dashed) vertical line on the left identifies the peak of mode B, and that on the right indicates twice the frequency of that on the left. The right panel shows the IF of modes B (top) and D (bottom). We used the ridge line to extract the IF because modes B and D are steep.  \label{fig:fourierSASI}}
\end{center}
\end{figure*}

The amplitude of mode B is $\sim$10 times larger than that of mode D. Because modes B and D seem almost static in the TFR, their fine structures are investigated. The left panels in Figure \ref{fig:fourierSASI} shows the GW signal in the Fourier space with a Hamming window with a width equal to the period when modes B and D dominate the TFR. As shown in the top left panel, the side lobe of the Hamming window is negligible compared with the peaks identified in the bottom left panel (see also Appendix A, where we use a boxcar window instead). In the bottom left panel, we show the Fourier component of the GW signals and find that mode D is approximately located at the overtone of mode B ($f_D \approx 2 f_B$); however, this relationship is not exact. The overtone frequency for the cross mode ($f_{D,\times}$) is close to twice that of the positive mode of mode B, $2 f_{B,+}$, rather than that of the cross mode, $2 f_{B,\times}$. The opposite case is also true: $f_{D, \times} \approx 2 f_{B,+}$, and $f_{D,+} \approx 2 f_{B,\times}$. This feature is also clear from the IFs of modes B and D, as shown in the right panels of Figure \ref{fig:fourierSASI}. These slight differences ($\Delta f \sim 5$ Hz) between $h_+$ and $h_-$ for both modes B and D are also interpreted as the relative phase change of $h_+$ and $h_-$ with a beat frequency of $f_b = \Delta f/2$. Its timescale is of the order of 0.1 s because $1/f_b = 0.4$ s. A polarization analysis can show the relative phase change as the change in the circular polarization. In fact, we find such a change of the Stokes $V$ parameters with a timescale of 0.1 s (\S \ref{ss:BD}).

\subsubsection{Core Motion of the Proto-Neutron Star \label{ss:coreo}}

\begin{figure}[htbp]
\begin{center}
  \includegraphics[bb=0 0 652 425,width=\linewidth]{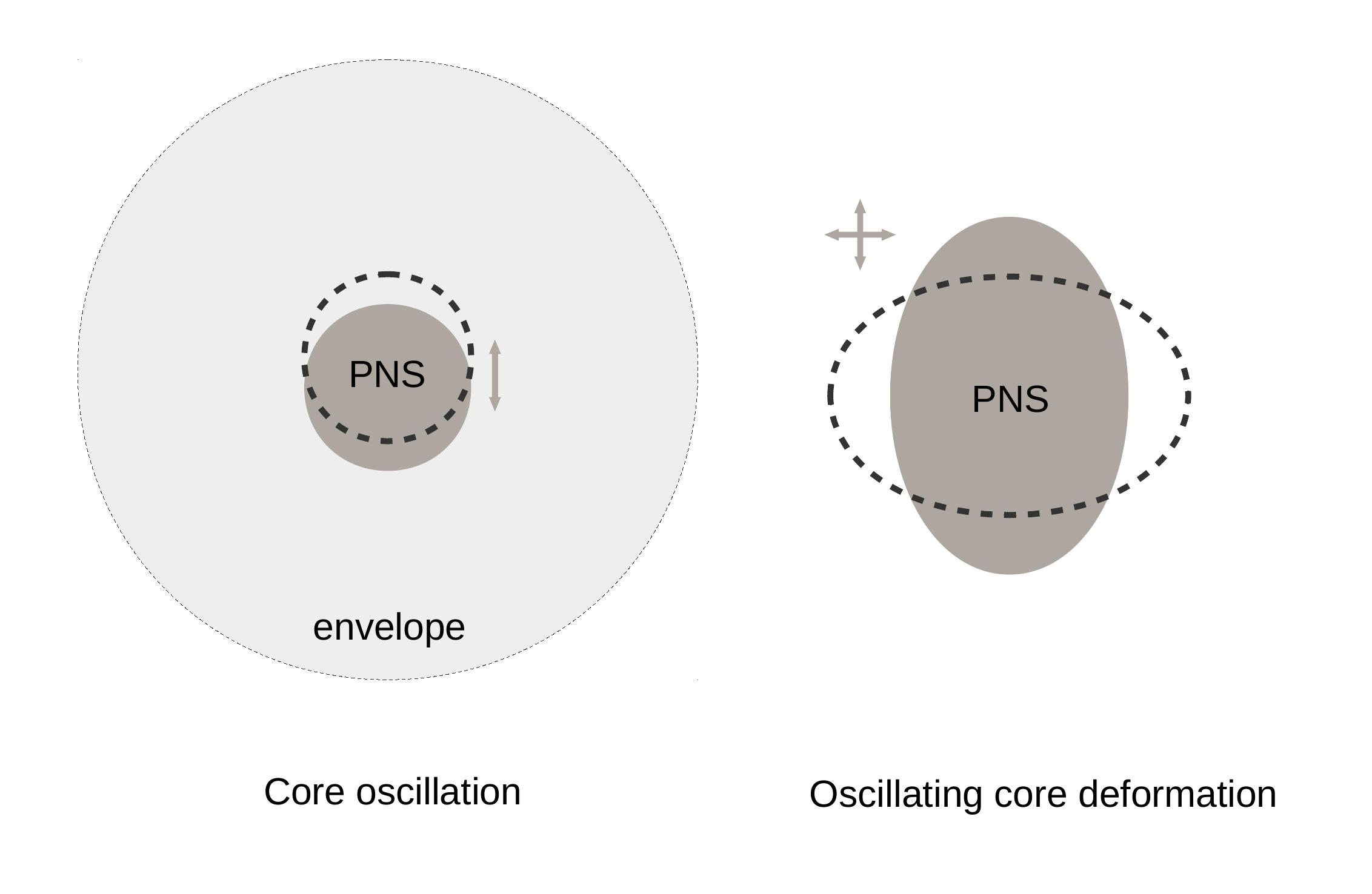}
  \caption{Schematic illustration of two possible models inducing modes B and D. \label{fig:schema}}
\end{center}
\end{figure}

To interpret the physical origin of the B and D modes, 
we consider two kinds of models that are schematically shown in Figure \ref{fig:schema}. The first is the core motion model (left panel) based on two harmonic oscillators that represent the core and ambient envelope. The core oscillates coherently in a direction and its momentum is transferred to the other oscillator, the ambient envelope. 
This core oscillation and/or rotation induce a change in the quadratic moment
that could be the origin of the two modes.
In this discussion we ignore the motion of the envelope for simplicity,
since the qualitative feature of the model is not affected by the motion.

Let us first suppose that the is core located at an offset from the center, $L$, 
and oscillates with a frequency $f=f_c$. 
The oscillation of a quadratic moment induced by this core oscillation provides oscillation frequencies of $f_c$ and $2f_c$. For an ideal example, if the core oscillates in the direction of the offset, we can write the motion of the center-of-mass (CoM)  as $r=A\cos(2\pi f t)+L$, where $A$ is the amplitude of the core-oscillation. The quadrupole moment of the core is easily calculated as $Mr^2=M(A^2\cos(2\pi (2 f_c) t) +2AL\cos(2\pi f_c t)+ {\rm const.})$, where $M$ is the mass of the core. There are two components of the oscillation with $f=f_c$ and $f=2f_c$, and these two modes may correspond to the B and D modes, respectively.

\begin{figure}[htbp]
\begin{center}
  \includegraphics[bb=0 0 562 399,width=\linewidth]{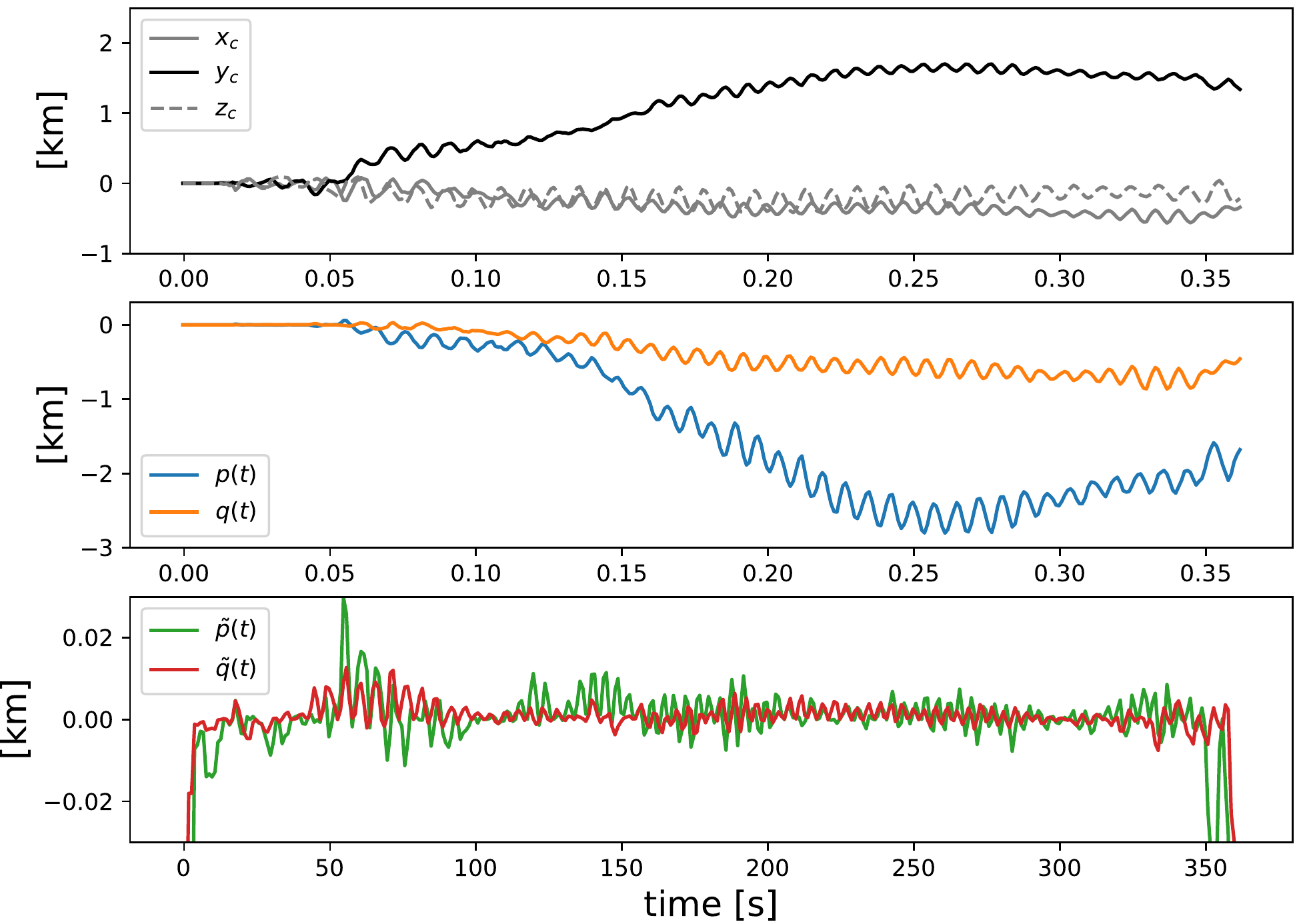}
  \caption{Top: Core motion of the PNS (the CoM of the PNS is defined for $\rho > 10^{13} \mathrm{g \, cm^{-3}}$). Middle: Quadratic components of the core motion, $p(t)$ and $q(t)$, toward the $z$ direction. Bottom: High-pass-filtered ($>$50 Hz) quadratic components, $\tilde{p}(t)$ and $\tilde{q}(t)$.  \label{fig:posq}}
\end{center}
\end{figure}

In a realistic case, the vector components should be taken into account.
The top panel of Figure \ref{fig:posq} shows the CoM position of the PNS defined for a density of $\rho > 10^{13} \mathrm{g \, cm^{3}}$ in our simulation.  
The position of the CoM of the PNS oscillates with a frequency of $\sim 130 $Hz, similar to that of the B mode.
Each component of CoM oscillates with an amplitude of the order of several kilometers. 
In addition, the drifting of the CoM of the PNS results in an offset from the CoM of the system, which is about 1--2 km toward the $y$ direction after $t=0.1$ s. We note that the CoM in the larger boxes converge to a constant value close to zero. 

Our simulated GWs are computed toward the $z$ direction; therefore, the $I_{xx}-I_{yy}$ and $I_{xy}$ components of the quadrupole moment tensor are involved in the GWs.
To calculate its contribution, we define two variables: 
\begin{eqnarray}
\label{eq:quv}
p(t) &\equiv& x_c^2(t) - y_c^2(t), \\
q(t) &\equiv& x_c(t) y_c (t), 
\end{eqnarray}
where $x_c(t)$ and $y_c(t)$ are the $x$ and $y$ components of the CoM of the PNS, respectively. The middle panel in Figure \ref{fig:posq} shows $p(t)$ and $q(t)$.
The 130 Hz oscillation occurs around an offset of $L\sim 2$ km from the CoM. The cross term between this oscillation and the offset induces a 130 Hz oscillation of the GW. 
The amplitude of the oscillation is $A \sim 0.3$ km. 

The order of the GW amplitude is estimated using the parameters obtained by the analysis above,
\begin{eqnarray}
\label{eq:orderco}
h &\sim& 2 G M A (2 L) (2 \pi f_c)^2 \nonumber \\
&=& \displaystyle{ 10^{-22} \frac{M}{M_\odot}\frac{L}{2 \mathrm{km}} \frac{A}{0.3 \mathrm{km}} \left(\frac{f_c}{130 \mathrm{Hz}} \right) ^2}\times
 \nonumber \\
&\,&
\displaystyle{\left( \frac{D}{10 \mathrm{kpc}} \right)^{-1} },
\end{eqnarray}
where $M$ is the mass of the oscillator (corresponding to the PNS mass), and $D$ is the distance of the GW source. The estimated value is comparable to the GW amplitude. Therefore, the static quadratic oscillation resulting from core oscillation with $f=130$ Hz is a possible origin of mode B. 

\begin{figure*}[htbp]
\begin{center}
  \includegraphics[bb=0 0 491 328,width=0.49\linewidth]{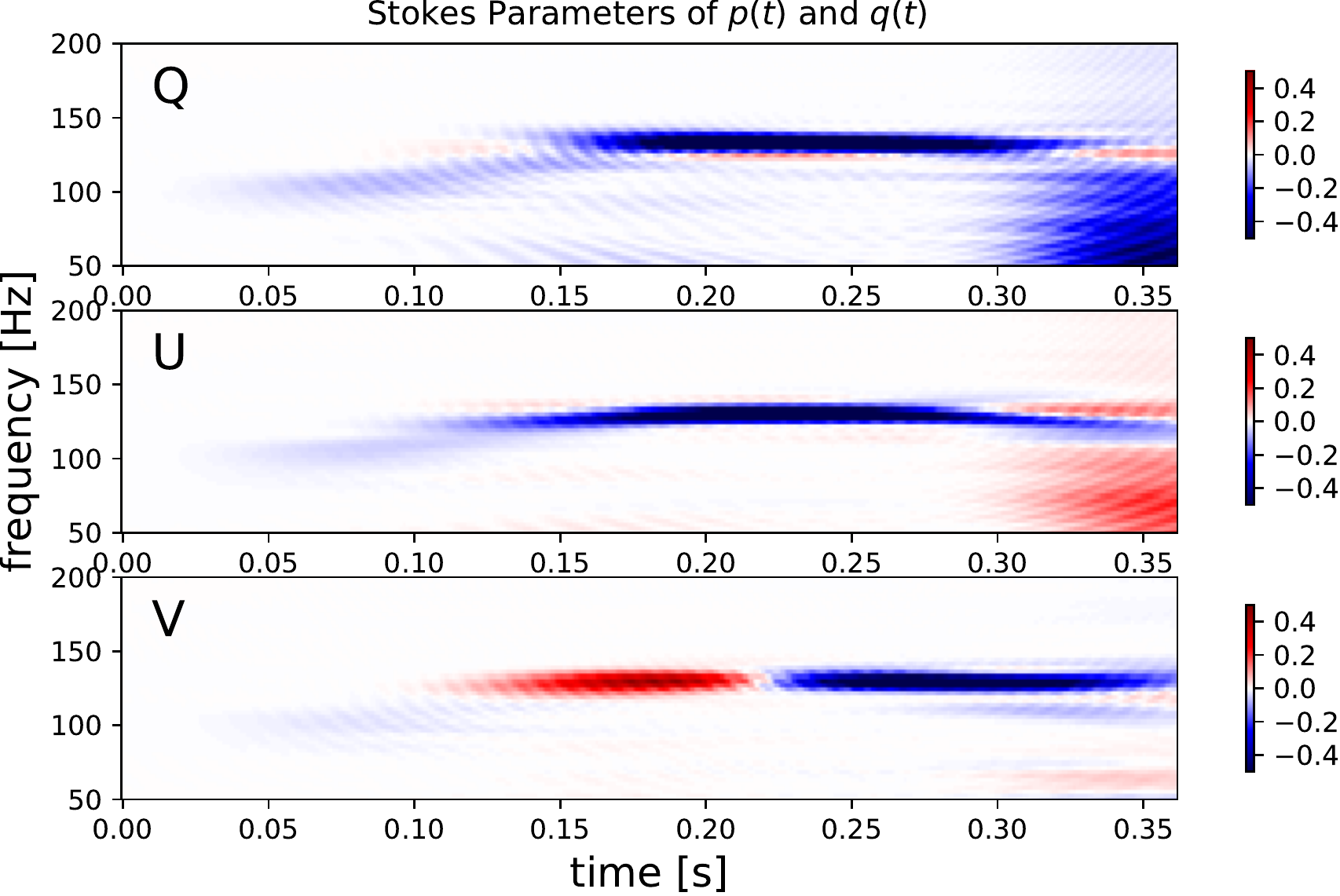}
  \includegraphics[bb=0 0 471 323,width=0.47\linewidth]{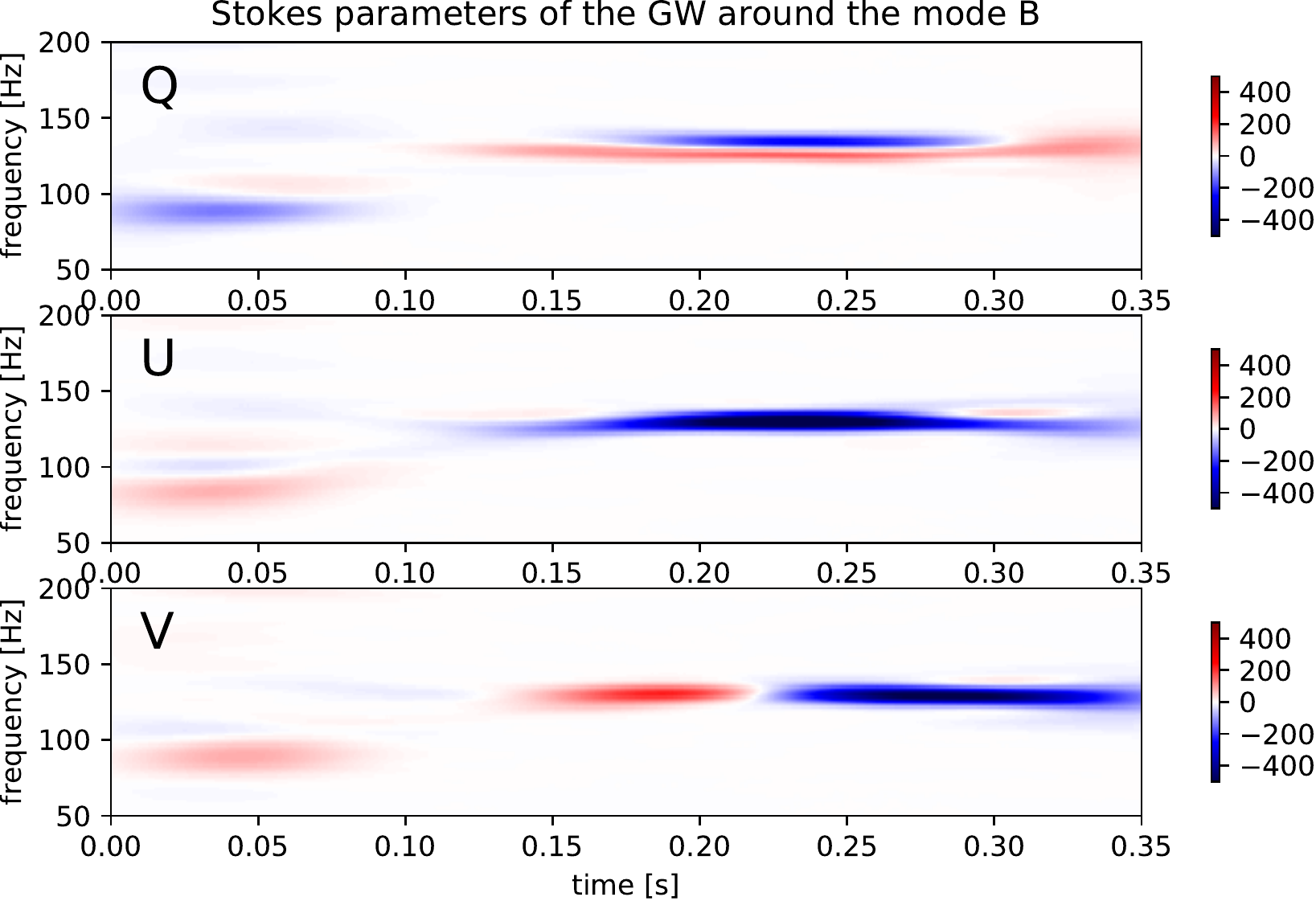}
  \caption{ Right: Stokes parameters of a displacement of the PNS ($\rho > 10^{13} \mathrm{g \, cm^{-3}}$) that is representative of the PNS core oscillation, $p(t),q(t)$. Left: Stokes parameters of the GWs around mode B. Both TFRs are of spectrogram type. \label{fig:quvoscB}}
\end{center}
\end{figure*}

The panels in Figure \ref{fig:quvoscB} show spectrogram-type Stokes parameters of $p$ and $q$ (left), and mode B (right). We found a similar time evolution of the polarization between $p$ and $q$, and the B mode. This correspondence also supports this simple modeling.

\begin{figure*}[htbp]
\begin{center}
  \includegraphics[bb=0 0 483 326,width=0.49\linewidth]{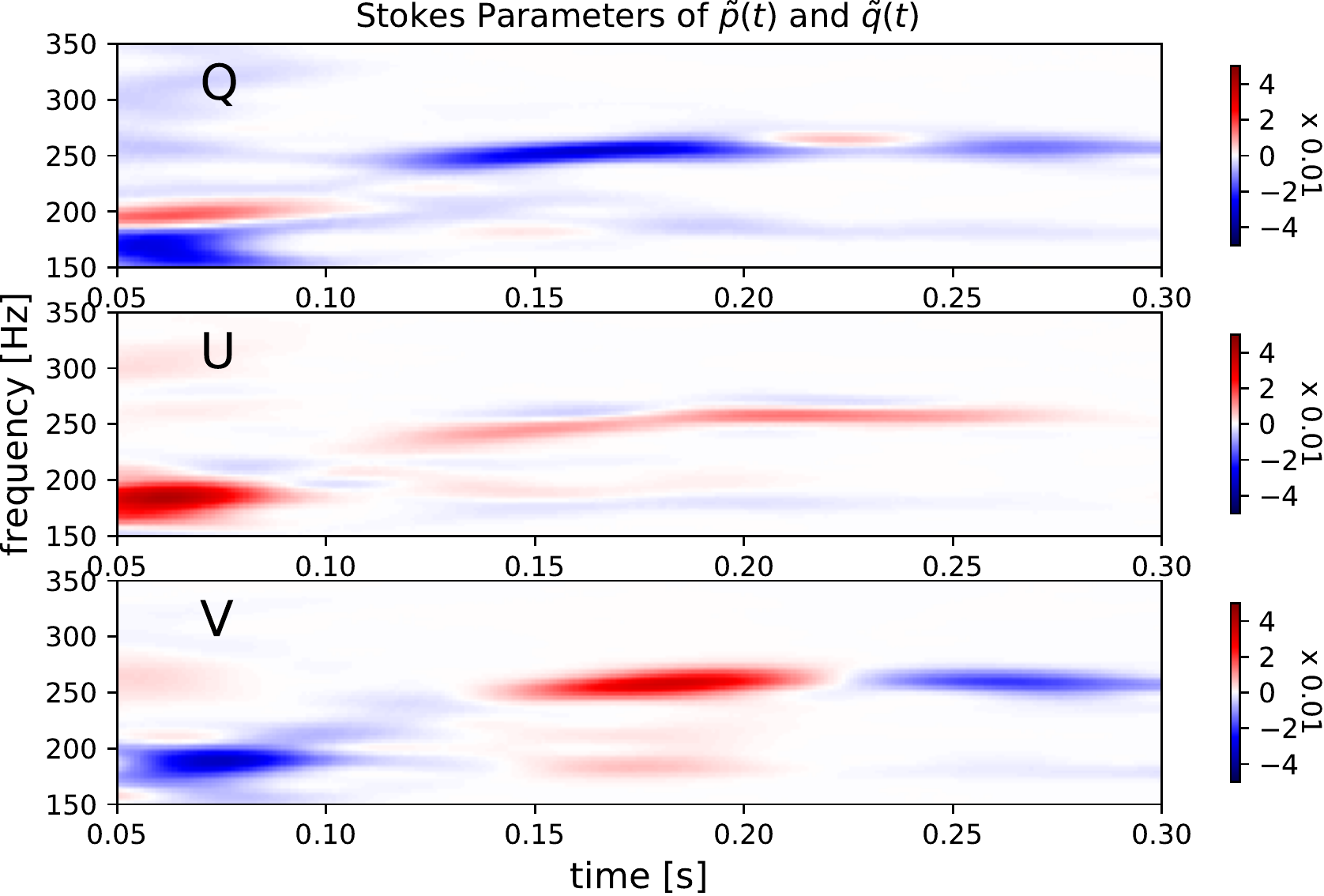}
  \includegraphics[bb=0 0 473 323,width=0.47\linewidth]{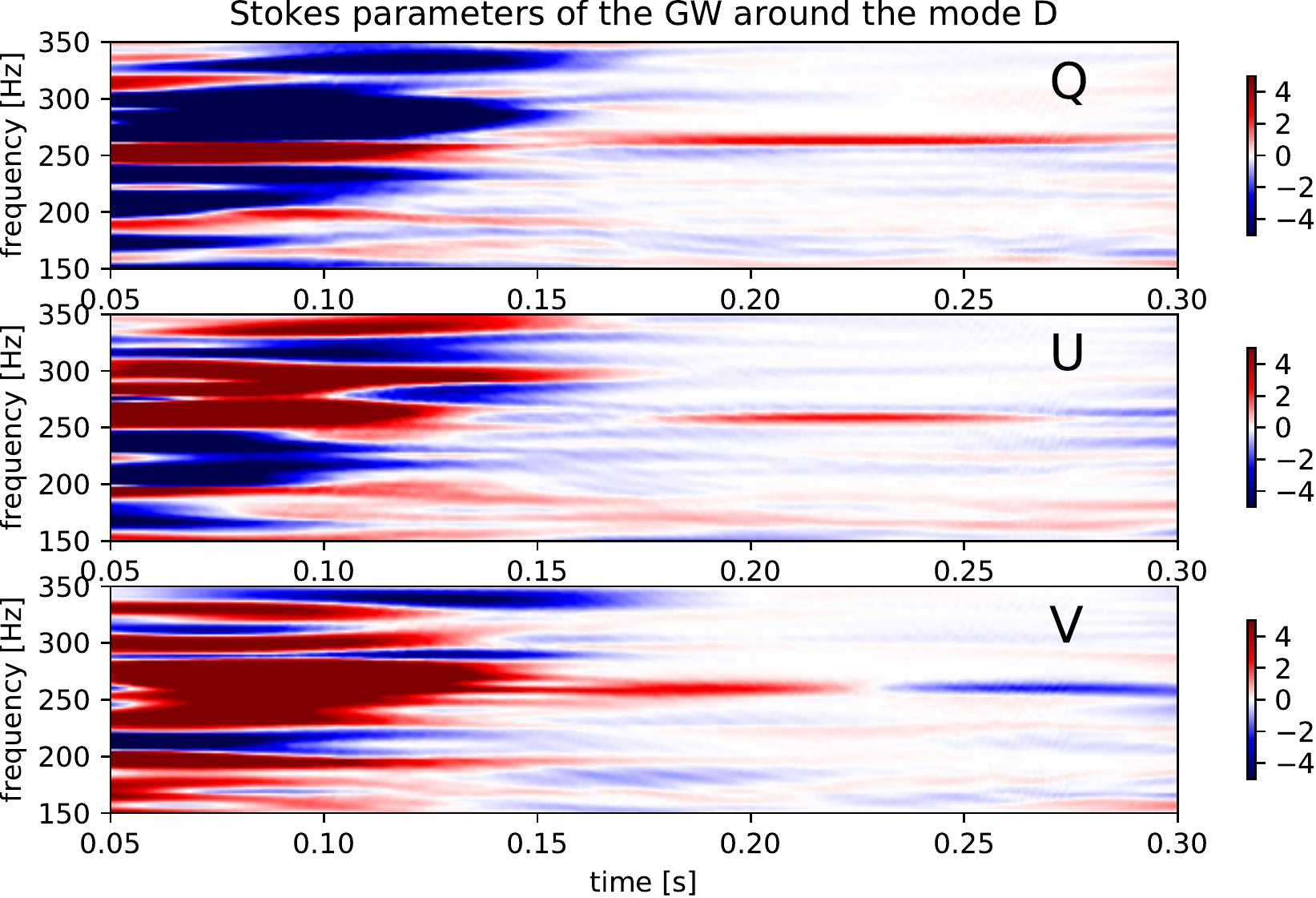}
  \caption{Left: Stokes parameters of a displacement of the PNS ($\rho > 10^{13} \mathrm{g \, cm^{^3}}$) after the high-pass filter, $\tilde{p}(t), \tilde{q}(t)$. Stokes parameters of GWs around mode D. Both TFRs are spectrogram type. \label{fig:quvoscD}}
\end{center}
\end{figure*}

The overtone of the core oscillation is a possible origin of mode D. To see this component, we apply a high-pass filter ($> 50$ Hz) to the core motion and define the quadratic moment as
\begin{eqnarray}
\label{eq:quv}
\tilde{p}(t) &\equiv& \tilde{x}_c^2(t) - \tilde{y}_c^2(t), \\
\tilde{q}(t) &\equiv& \tilde{x}_c(t) \tilde{y}_c (t), 
\end{eqnarray}
where $\tilde{x}(t)$ and $\tilde{y}(t)$ are the high-pass-filtered ($>$50 Hz) displacements of the CoM of the PNS. As shown in the bottom panel in Figure \ref{fig:posq} ($\tilde{p}(t)$ and $\tilde{q}(t)$) and the left panels in Figure \ref{fig:quvoscD} (spectrograms), the frequencies of $\tilde{p}(t)$ and $\tilde{q}(t)$ are exactly twice those of $p(t)$ and $q(t)$, respectively, and their changes in polarization are closely related.

By analogy with the harmonic oscillator \citep[e.g.,][]{maggiore2008gravitational}, the GW amplitude of the overtone of the core oscillation can be estimated as
\begin{eqnarray}
\label{eq:ordercohigh}
h &\sim& 2 G M A^2 (2 \pi f_c)^2 \nonumber \\
&=& \displaystyle{ 10^{-23}  \frac{M}{M_\odot} \left(\frac{A}{0.3 \mathrm{km}} \right)^2 \left(\frac{f_c}{130 \mathrm{Hz}} \right) ^2 \left( \frac{D}{10 \mathrm{kpc}} \right)^{-1} }.
\end{eqnarray}
The amplitude of the overtone (260 Hz) is an order of magnitude smaller than that of the fundamental mode, 130 Hz, which is consistent with the ratio of mode B to mode D.

The right panels in Figures \ref{fig:quvoscB} and \ref{fig:quvoscD} show the (spectrogram-type) Stokes parameters around modes B and D, respectively. Comparing these with the left panels, there are significant similarities, in particular, the point at which the circular polarization $V$ changes, $t \sim 0.22$ s. This similarity also supports the core motion origin of modes B and D.

\subsection{Oscillating Quadratic Deformation of the Proto-Neutron Star Core}

\begin{figure}[htbp]
\begin{center}
\includegraphics[bb=0 0 334 200,width=\linewidth]{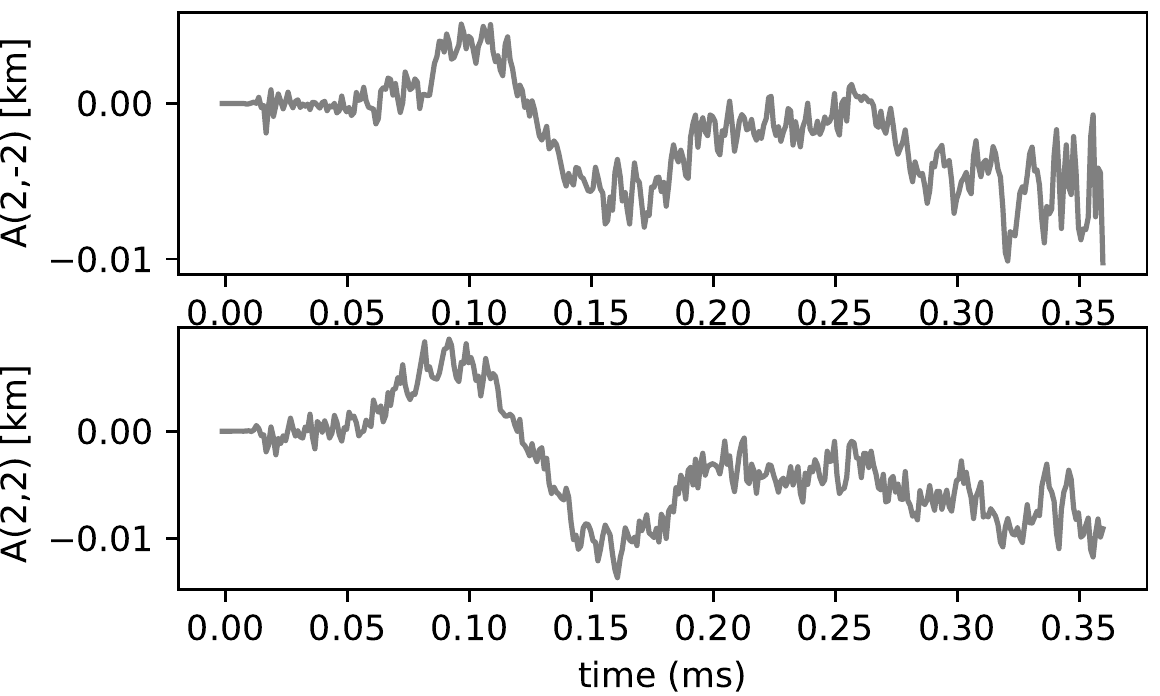}
\caption{Quadratic components of the isodensity surface of the PNS ($\rho = 10^{14} \, \mathrm{g \, cm^{-3}}$) after the coherent motion of the PNS is removed. We choose $A_{2,2}$ and $A_{2,-2}$ corresponding to GWs propagating toward the $z$ direction. \label{fig:Alm2}}
\end{center}
\end{figure}

\begin{figure*}[htbp]
\begin{center}
\includegraphics[bb=0 0 484 323,width=0.49\linewidth]{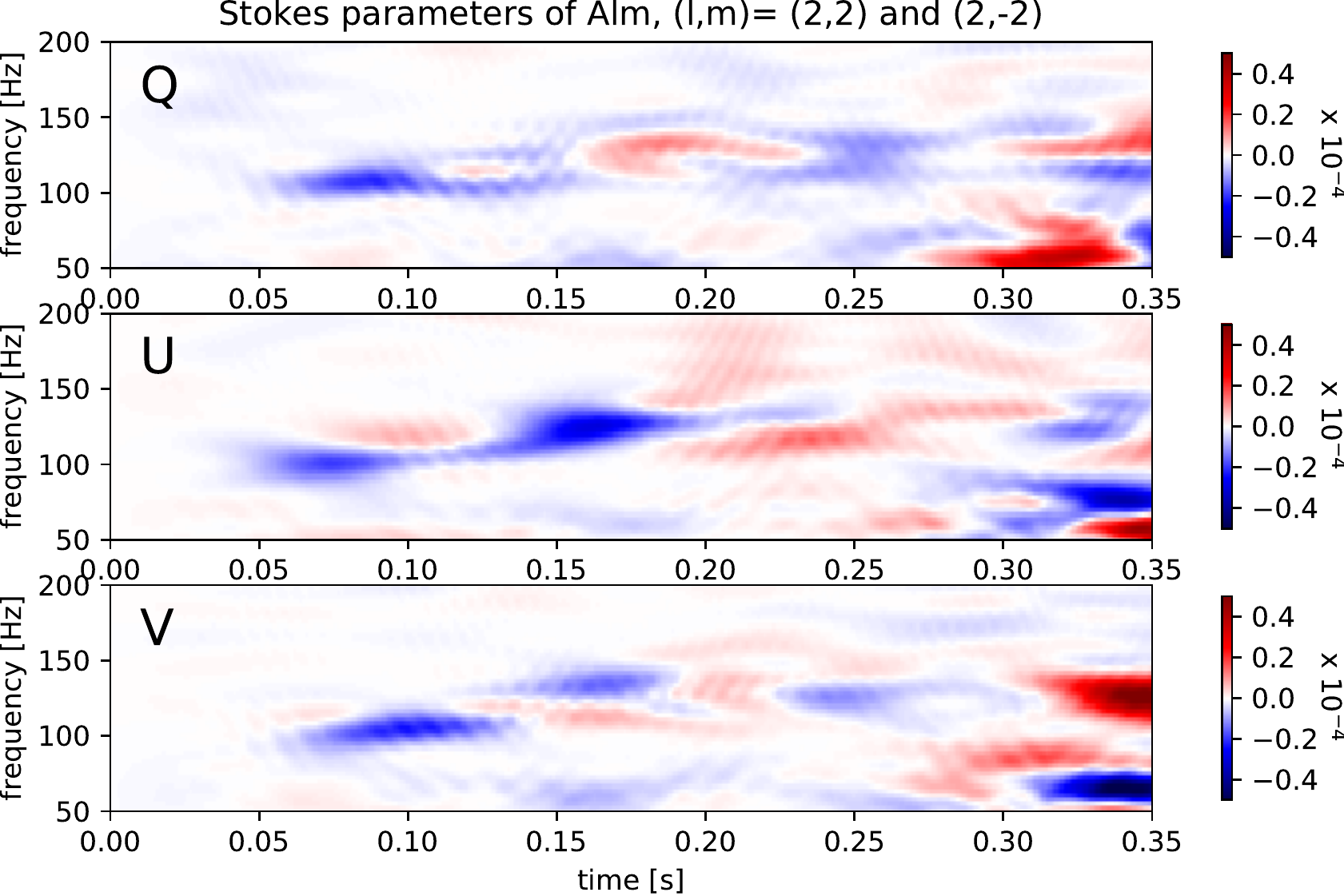}
  \includegraphics[bb=0 0 473 323,width=0.47\linewidth]{spstokesSFHx_4_B.pdf}
\caption{Comparison of Stokes parameters of $A_{2,\pm 2}$ of the isodensity of the PNS shown in Figure \ref{fig:Alm2} (left) with the Stokes parameters of the GW around mode B (right, same as \ref{fig:quvoscB}). Both TFRs are spectrogram type. \label{fig:StokesAlm2}}
\end{center}
\end{figure*}

Another possible origin of modes B and D is the oscillating quadratic deformation of the PNS, as shown in the right panel of Figure \ref{fig:schema}. 
To quantitatively observe the core surface deformation, 
the time evolutions of the mode amplitudes, $A_{lm}$, of the spherical polar expansion of the isodensity surface $R_{14}(\theta,\phi)$ are plotted (see \cite{burrows12} for the definition of the mode amplitudes of a 2-sphere).
The isodensity surface at the rest mass density of $\rho=10^{14}$ g cm$^{-3}$ is exracted.
Here $(l,m)$ is the polynomial degrees, and $\theta$ and $\phi$ denote the polar and azimuth angles, respectively.
The weak oscillation of the quadratic components of the isosurface density after the coherent motion of the PNS around 130 Hz is removed, as shown in Figure \ref{fig:Alm2}. The amplitude is roughly $C \lesssim 3 \times 10^{-3}$ km. The estimated GW amplitude from this oscillating deformation is
\begin{eqnarray}
\label{eq:qosc}
h &\sim& 2 G M C R (2 \pi f_c)^2 \nonumber \\
&=& \displaystyle{ 10^{-23}  \frac{M}{M_\odot} \left(\frac{C}{3 \times 10^{-3} \mathrm{km}} \right)  } \nonumber \\
&\,&\displaystyle{  \left(\frac{R}{30 \mathrm{km}} \right) \left(\frac{f_c}{130 \mathrm{Hz}} \right) ^2 \left( \frac{D}{10 \mathrm{kpc}} \right)^{-1} },
\end{eqnarray}
where $R$ is the core radius. This value is an order smaller than the amplitude expected from the core oscillation according to equation (\ref{eq:orderco}). Figure \ref{fig:StokesAlm2} compares the spectrogram-type Stokes parameters of $A_{2,\pm 2}$ and mode B of Figure \ref{fig:allstokes}, and there are no significant similarities. These results suggest that the oscillating quadratic deformation of the PNS core represents a relatively minor contribution to mode B.

\section{Quadratic Time-Frequency Analysis of Noisy Data}\label{sec:noise}

The aforementioned analysis of the waveform in the absence of noise demonstrated the useful features of the quadratic TFRs for understanding the nature of the simulated GWs. A further question is whether the quadratic TFRs are applicable to future real GW data.　In this section, we test the quadratic TFRs and their Stokes representations under realistic noise, to determine their ability to identify real GW signals, by a coherent network analysis \citep{2018MNRAS.477L..96H}. 

\begin{figure*}[htbp]
\begin{center}
  \includegraphics[bb=0 0 410 285,width=\linewidth]{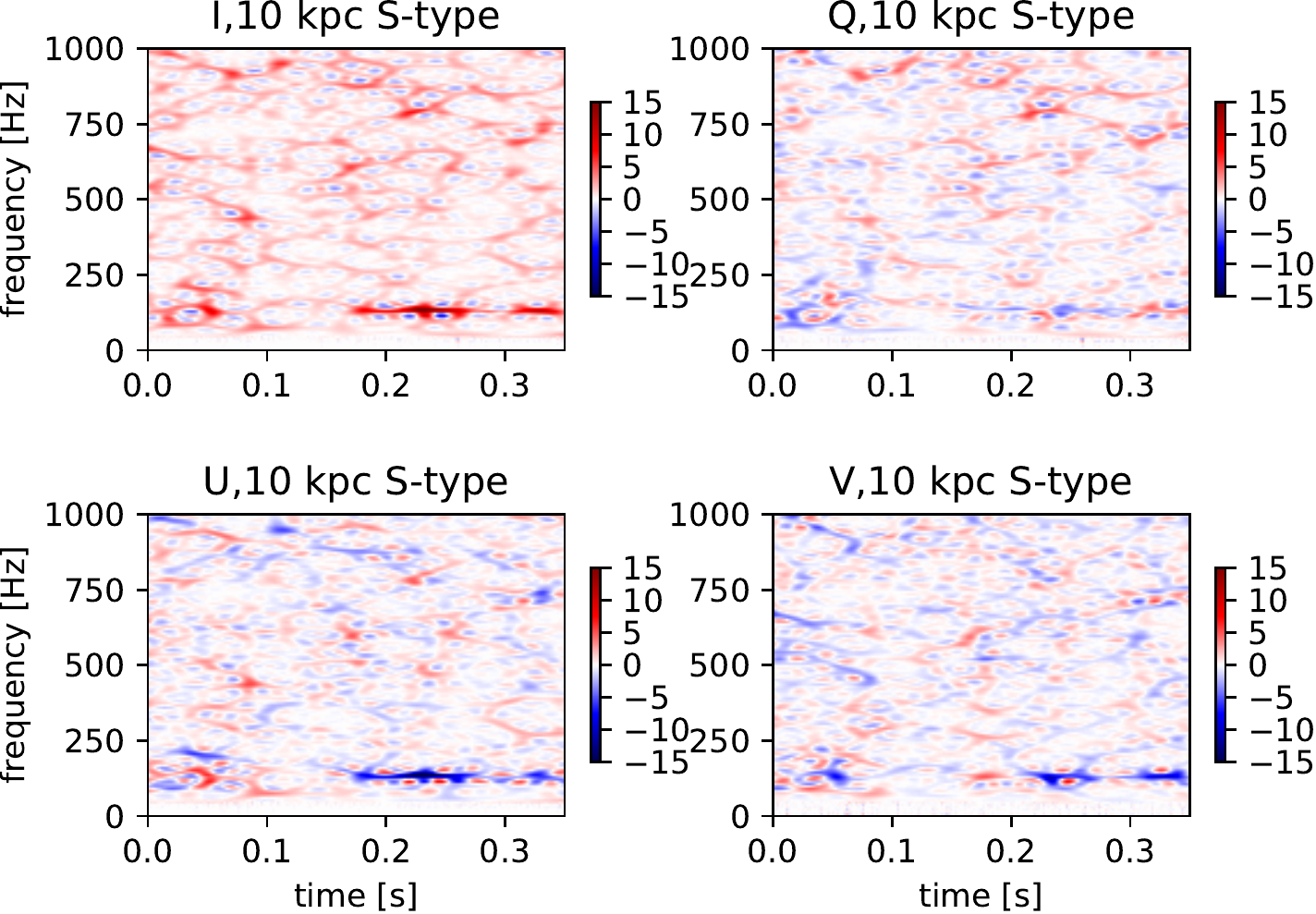}
  \caption{The S-type Stokes parameters for the simulated waveforms with noise assuming the CCSN is 10 kpc away from the four detectors  \citep[LIGO Hanford, LIGO Livingstone, VIRGO and KAGRA; ][]{2018MNRAS.477L..96H}.  \label{fig:compStokes}}
\end{center}
\end{figure*}

\begin{figure}[htbp]
\begin{center}
  \includegraphics[bb=0 0 327 200,width=\linewidth]{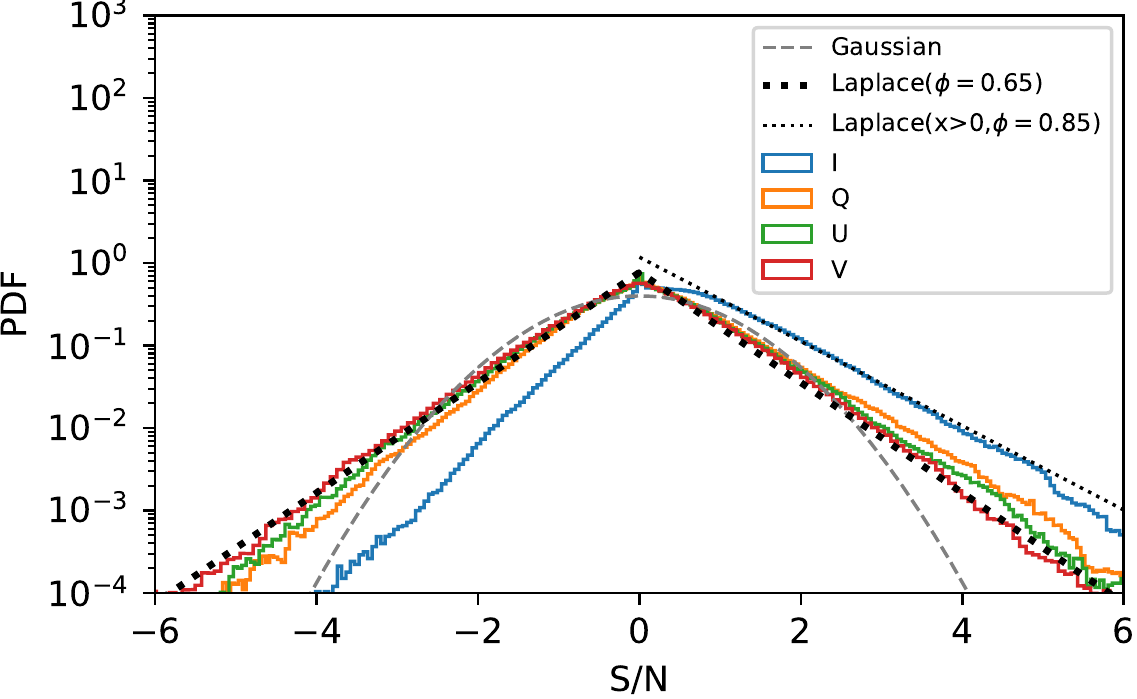}
  \caption{The distributions of the S/N for the $I, Q, U,$ and $V$ maps. The dashed and dotted lines indicate a normal distribution and Laplace distribution with $\phi=0.65$, respectively. \label{fig:sm_dist}}
\end{center}
\end{figure}

Figure \ref{fig:compStokes} shows the S-type Stokes parameters for the GW signal from $d=10$ kpc with simulated noises assuming a combined signal by LIGO Hanford, LIGO Livingstone, VIRGO, and KAGRA \citep{2018MNRAS.477L..96H}. In this Figure, we use the TFR whitened by the power spectrum of the noise-only data. We find a clear signal of mode B, a barely detectable signal from mode A at $t \sim$ 0.35 [s], and at the beginning of the modes ($t \sim 0.05$ [s]). In the TFR of the $V$ parameter, the changing point of polarization at $t \sim 0.2$ [s] can be marginally seen. Although we refer the values in Figure \ref{fig:compStokes} to the S/N (signal to noise ratio), the probability density distributions (PDFs) are far from Gaussian as shown in Figure \ref{fig:sm_dist}. The PDFs of the $Q, U,$ and $V$ parameters are close to Laplace distributions with $\phi=0.65$,
\begin{eqnarray}
p(x) = \frac{1}{2 \phi} \exp{\frac{-|x|}{|\phi|}}.
\end{eqnarray}
The S/Ns corresponding to $p=0.0024$ (3 $\sigma$ in Gaussian) is S/N=3.5 and $p=5 \times 10^{-7}$ (5 $\sigma$ in Gaussian) is S/N = 9, after applying the Laplace distribution as the PDF of the $Q,U$, and $V$ parameters. 

\citet{2018MNRAS.477L..96H} suggested that the circular polarization improves the detectability of the GWs from CCSNe compared with the intensity. We also confirm that the $I$ parameters exhibits more noises than those in other Stokes parameters. Because the signals are a similar level in the S/N for all of the Stokes parameters, this fact is attributed to the difference of the PDF, i.e., the PDF of the I parameter has a longer tail (a shallower gradient) and is defined as a positive value (although there are leakages to the negative side due to noise), while other Stokes parameters can be both positive and negative. A positive definite Laplace distribution with $\phi=0.85$ is well fitted to the tail of the PDF of the $I$ parameters as shown by the thin dotted line in Figure \ref{fig:sm_dist}. This indicates that there is more noise for high S/Ns in the TFR of the $I$ parameter, compared with the other Stokes parameters. We also confirm that the use of spectrogram-type TFRs instead of S-type TFRs does not qualitatively change the results. 

\section{Summary and Discussion}\label{sec:summary}

In this paper, we applied the quadratic TFR and spectrogram analysis of the waveform in \citet{2016ApJ...829L..14K}. We summarize our findings as follows:
\begin{itemize}
\item The high-resolution TFR obtained by the S-method is useful for identifying modes in complex multimodal GWs from CCSNe. Using the S-method, we found that the quasistatic mode consists of a fundamental mode and its overtone, and also identified a new mode, mode C, which was overlooked in the previous analysis.  
\item The TFR of the Stokes parameters is a useful probe for exploring the physical origins of each mode. A random change in the polarization state may imply a convection-induced GW (modes A and C). The quasistatic polarization state in modes B and D is likely to imply that the dynamical oscillation of the PNS induces those modes. 
\item The polarization feature of mode B is detectable even for a combined signal from a CCSN at 10 kpc by LIGO Hanford, Livingstone, VIRGO, and KAGRA. 
\end{itemize}

By utilizing {\color{red} the spectrogram with the adequate window shape and size and also the S-method, we could more clearly identify several distinct features hidden in the gravitational waveforms, which were overlooked in the previous analysis \citep{2016ApJ...829L..14K}.}
Moreover, compared with the spectrogram analysis that is currently widely used in the CCSN study, we can specify the instantaneous peak frequency of each component.
This may lead to the next stage of clearly understanding CCSN dynamics from future GW detections, as 
many of the oscillation modes can be excited depending on the PNS structure in the PNS core.
Each mode has its characteristic frequency and potentially leads to GW emissions at the relevant frequency.
Therefore, by using asteroseismology of CCSNe \citep{torres18,viktoriya18}, the sharply determined frequency in the gravitational waveforms by the S-method can be connected to the excited mode in the PNS.
This may eventually lead the PNS structure being resolved.
 
The fact that the dominant frequency of the SASI is $f_\mathrm{SASI} \sim$ 65 Hz \citep{2016ApJ...829L..14K} hints at a connection between mode B and the SASI because this is half the frequency of mode B. It is unlikely that the dipole oscillation of the SASI induces the core oscillation of the PNS,  discussed in \S \ref{ss:coreo}, because there is a frequency mismatch. A possible explanation is that the overtone ($f=$ 130 Hz) of the quadratic components of the SASI oscillation is related to the core oscillation and oscillation deformation. 

It is worth mentioning that \cite{yoshida07} investigated the excitation mechanism of the PNS oscillation induced by the SASI. 
Their linear analysis showed that if there is a significant perturbation on the PNS surface ($r\sim50$ km), e.g., by the SASI, $g$-modes and its higher overtones can be indeed excited at $r\sim$10 and 30 km.
According to \cite{2016ApJ...829L..14K}, the spiral SASI mode becomes active at $t \ga 200$ ms in model SFHx without a fixed orientation. The impact of the spiral SASI on exciting the PNS eigen-modes has yet to be studied.
Given the complexity of the non-linear mode coupling 
(e.g., \citet{weinberg}),
it is not straightforward to connect the SASI modes that take place at $r\sim100$ km and the subsequent mode excitations at $r\la50$ km. We are now attempting to perform a more detailed linear perturbation analysis of the PNS core in a fully relativistic framework to find the excitable modes \citep[see,][]{sotani17}. Because this paper focuses on the methodology of the TFA, this is the subject of future work (Sotani et al. in prep.) 

The TFR package used in this paper is publicly available via github under the GNU General Public License \footnote{https://github.com/HajimeKawahara/juwvid}.

H.K. is supported by a Grant-in-Aid for Young Scientists (B) from JSPS (Japan Society for the Promotion of Science)(No.\,JP17K14246) and JSPS KAKENHI (Nos. JP18H01247 and JP18H04577). 
T.K. is supported by the European Research Council (ERC; FP7) under ERC Advanced Grant FISH-321263 and ERC Starting Grant EUROPIUM-677912.
This work was partially supported by the JSPS Core-to-Core Program Planet$^2$. This study was also supported by JSPS KAKENHI (Nos. JP15H00789, JP15H01039, JP17H01130, and JP17H06364), the Central Research Institute of 
Fukuoka University (Nos. 171042 and 177103), and JICFuS as a priority issue to be tackled using a Post `K' Computer.

\appendix
\section{A. Aliasing effect of the window}

\begin{figure}[htbp]
\begin{center}
\includegraphics[bb=0 0 504 360,width=0.51\linewidth]{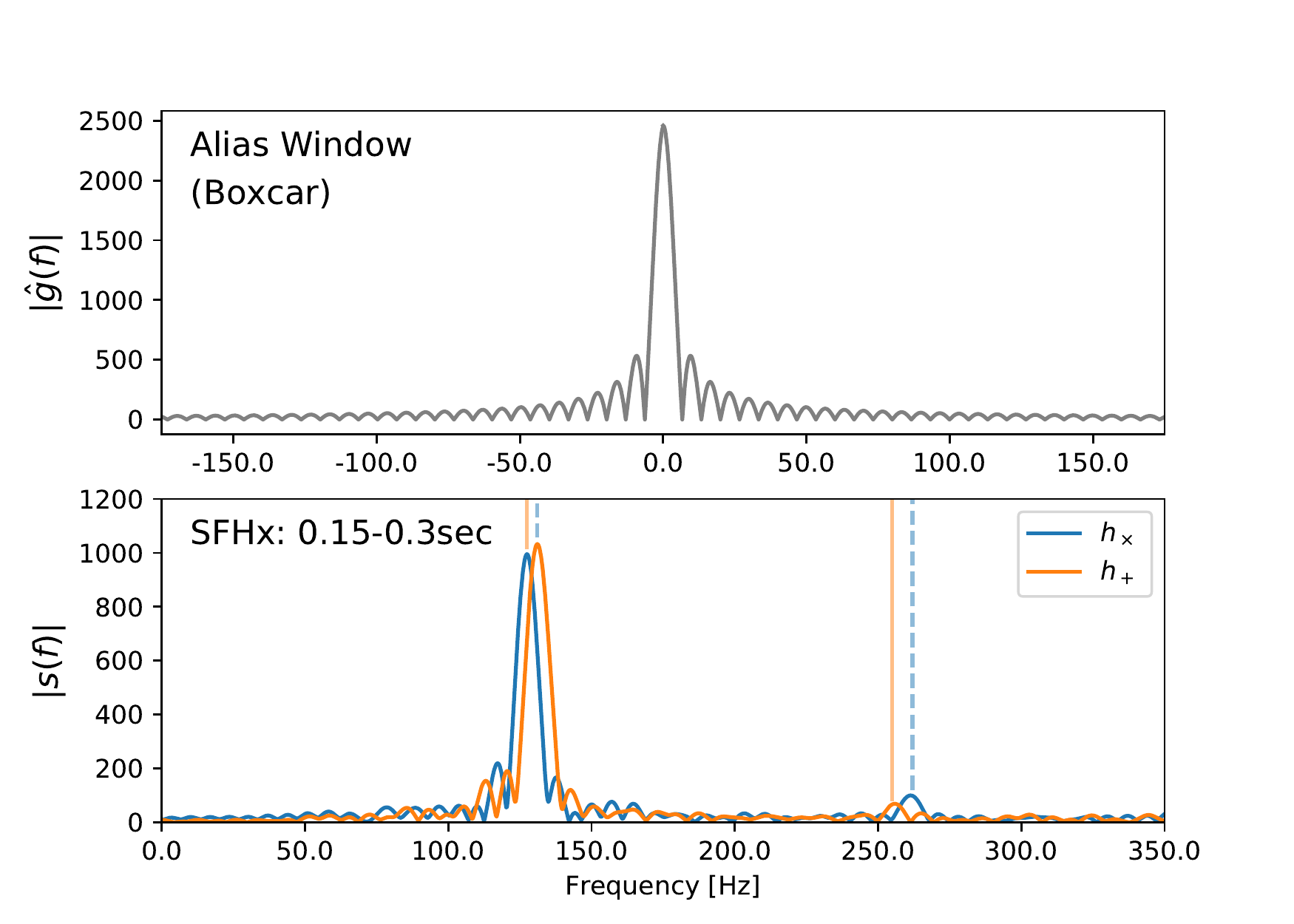}
\caption{ The same plot as the left panel in Figure \ref{fig:fourierSASI}, except for a boxcar window (instead of the Hamming window). \label{fig:alias}}
\end{center}
\end{figure}

Denoting the Fourier transform of $F$ by $\hat{F}$, i.e., $\hat{F}(\fif) = \int_{-\infty}^\infty F(\tau) e^{- 2 \pi i \fif \tau} d\tau$, the windowed Fourier transform for a symmetric window can be expressed as
\begin{eqnarray}
\label{eq:stftdef3}
s(\fif, t) &=& \int_{-\infty}^\infty y(\tau) g(t - \tau,l) e^{- 2 \pi i \fif \tau} d\tau, \\
&=& \int_{-\infty}^{\infty} \hat{y}(\fif - \fif^\prime) \hat{G}(\fif^\prime,t)  d\fif^\prime  = \int_{-\infty}^{\infty} e^{-2 \pi i \fif^\prime t} \hat{y}(\fif - \fif^\prime) \hat{g}(\fif^\prime,t) d\fif^\prime 
\end{eqnarray}
where we use
\begin{eqnarray}
\label{eq:stftdef2}
\hat{G}(\fif, t) = \int_{-\infty}^\infty  g(t - \tau,l) e^{- 2 \pi i \fif \tau} d\tau = e^{-2 \pi i \fif t} \int_{-\infty}^\infty  g(-\tau^\prime,l) e^{-2 \pi i \fif \tau^\prime} d\tau^\prime = e^{-2 \pi i \fif t} \hat{g}(\fif,t),
\end{eqnarray}
assuming a symmetric window, $g(-t,l)=g(t,l)$. Thus, the shape of the Fourier transform of the window $\hat{g} (\fif)$ is convolved into the windowed Fourier transform of the signal. The Fourier component of the Hamming window, as shown in the top left panel in Figure \ref{fig:fourierSASI}, suffers less from the issue of side lobes compared with $\hat{g} (\fif)$. Figure \ref{fig:alias} displays the same plot as the left panel in Figure \ref{fig:fourierSASI}, but for a simple boxcar window. In this case, there are severe side lobe effects around the real peaks. This aliasing effect may cause artifacts in the TFRs. 

\end{document}